\documentclass[showpacs,preprintnumbers,amsmath,amssymb,floatfix,elsart,prd]{revtex4}

\usepackage{graphicx}
\usepackage{dcolumn}
\usepackage{bm}

\begin{document}

\title{Generating functions of wormholes}

\author{Farook Rahaman}
\email{rahaman@associates.iucaa.in} \affiliation{Department of
Mathematics, Jadavpur University, Kolkata 700 032, West Bengal,
India}

\author{Susmita Sarkar}
\email{susmita.mathju@gmail.com} \affiliation{Department of
Mathematics, Jadavpur University, Kolkata 700 032, West Bengal,
India}

\author{Ksh. Newton Singh}
\email{ntnphy@gmail.com } \affiliation{Faculty of Sciences, Department of Physics, National Defence Academy,
 Khadakwasla,   Pune, India \\
 and \\
 Department of
Mathematics, Jadavpur University, Kolkata 700032, West Bengal,
India}

\author{Neeraj Pant}
\email{neeraj.pant@yahoo.com} \affiliation{Faculty of Computational Sciences, Department of Mathematics, National Defence Academy, Khadakwasla, Pune-411023, India.}

\date{\today}

\begin{abstract}
It is known that   wormhole geometry could be found solving the Einstein field equations by tolerating the violation of null energy condition (NEC). Violation of  NEC is not possible for the physical matter distributions, however, can be achieved by considering distributions of "exotic matter". The main purpose of this work is to find generating functions comprising the wormhole like geometry and discuss the nature of these generating functions. We have used the Herrera et al. \cite{1} approaches of obtaining generating functions in the background of wormhole spacetime. Here we have adopt two approaches of solving the field equations to find wormhole geometry. In the first method, we have assumed the redshift function $f(r)$ and the shape function $b(r)$ and solve for the generating functions. In an another attempt we assume generating functions and redshift functions  and then try to find  shape functions of the wormholes.\\
\\

\end{abstract}

 \maketitle

\section{Introduction}

In recent time, the investigation of  wormhole solutions has been an stimulating issue of discussion in astrophysical point of view. The origin of wormhole can be tracked back to 1916 where Flamm \cite{FL} sketches the equatorial plane of the spatial sections of the Schwarzschild's interior solution. Further analysis reveals that the surface of revolution is isometric to a planar section of the Schwarzschild exterior solution. He also found that the meridional curve is a parabola, where the surface of revolution joins two asymptotically flat sheets. This concept is equivalent to the modern terminology of a tunnel or a wormhole geometry.\\

In 1935, Einstein and Rosen \cite{ER}  represented a wormhole-type solution to model elementary particles represented by a ``bridge" (now known as Einstein-Rosen Bridge) that connects two identical sheets. In other words, Einstein-Rosen bridge is an observational aspect to remove the Schwarzschild's coordinate singularity at $r=2M$ by a suitable coordinate transformation. Indeed, Einstein and Rosen found that certain coordinate transformations naturally cover only two asymptotically flat spacetimes of the maximally extended Schwarzschild spacetime. Thus, constructing a bridge demands the existence of an event horizon and a coordinate transformation avoiding singularity to artifact the Einstein-Rosen bridge. The eagerness of John Wheeler \cite{WH}  on topological issues in General Relativity reborn the concepts on wormhole only in 1955 and was subtle for two decades. Wheeler constructed ``geon" solutions by accounting coupled Einstein-Maxwell field equations in multiply connected spacetime, where two separated regions are connected by a tunnel. Geon is hypothetical ``unstable gravitational-electromagnetic quasi-soliton" \cite{VL}. In 1957, Misner and Wheeler \cite{MW}  used nontrival topology in source free Maxwell equations coupled with Einstein gravity to described electric charge classically and all other particle-like entities. This is also the same article where they first coined the term ``Wormhole". In another article, Wheeler \cite{whe} also described classical gravitation, electromagnetism, charge and mass in terms of curved empty space. Bronnikov \cite{bro1} introduced a scalar charge in scalar-tensor theory and proposed that by considering scalar charge as characteristics of elementary particles can obtained a different values of gravitational constant $G$. With the concept of wormholes and/or multiply connected topologies, arises a question of causality. If two signals are sending between two points where one traveling along longer route and other on shorter route through wormhole seems to violate causality condition. However, Fuller and Wheeler shown that in such condition the causality is preserved \cite{ful}. \\

In the meanwhile, a concept of ``drainhole" was introduced by Ellis \cite{ell} in 1973. A drainhole is static, spherically symmetric, horizonless and geodesically complete topological hole in spacetime manifold which can be form by coupling of a scalar field with the geometry of spacetime. G. Clement \cite{cle1} investigated wormhole solutions in higher dimensions. It is also shown that the field equations of sourceless in $n+p-$dimensions reduce to $n-$dimensions field equations coupled to a repulsive scalar field. In his second paper \cite{cle2} an axisymmetric regular multiwormhole was presented in five dimensions and interpreted multiwormhole solutions as multiparticle systems. In another paper, Clement \cite{cle3} shown that a weak perfect fluid source generates a small mass for the Kaluza-Klein wormhole without affecting the spacial geometry or the scalar field which remains short range. Interestingly, a regular and localized solution of field equations where the gravitational and electromagnetic fields are coupled with scalar field, can be interpreted as an extended charged particle \cite{cle4}.\\

However, only in 1988 the pioneering work of Morris and Thorne \cite{MT} in their seminal paper renaissance the physics of wormhole by proposing the traversable wormhole. This concept was completely  different  from the Einstein-Rosen bridge \cite{ER}. Wormholes are hypothetical routes through spacetime i.e. it acts as a tunnel which  connects two different regions of the same spacetime or different spacetimes \cite{ML}. A traversable wormhole is an exact  solution of Einstein field equations that permits  the viewers  to travel them  from one space to the other. The throat of the wormhole is defined by the  surface of minimal area connecting   two regions. According to Morris and Thorne \cite{MT}  the matter distribution that retains the throat open is  not real    i.e. it violates  the null energy condition (NEC:$\rho+P_r<0$). In other words,   the normal matter is incompetent to hold a wormhole open. For the last few decades, many theoretical physicists  search for exact wormhole solutions within the framework of general theory of relativity \cite{8,9,10,11,12,13,14,15,16,17,18,19,20,21,22}. Usually, physicists describe the wormhole structure in two ways:  either one first designs  the metric with the preferred  configurations  of a traversable wormhole, and then recuperated  the matter fields via Einstein's equation or assuming the matter distribution part to recover spacetime geometry  through  Einstein's equation. However,  if one  knows certain quantities of geometry and certain quantities of matter distribution, then one is capable to  find all the unknown  physical parameters of the wormhole structure  via Einstein field equations. Many other models of wormholes are also presented by several authors. The existence of regular solutions of the field equations with a non-linear sigma model as source is impossible without a center (as wormhole, horns, flux tubes) with positive-definite $h_{ab}(\Phi)$ and any potentials $V(\Phi)$ \cite{bro5}. Wormhole solutions when field equations minimally coupled with scalar field $\Phi$ revealed that (i) trapped-ghost wormholes are possible only with $V(\Phi)\neq 0$ (ii) wormholes with two asymptotically flat regions, a nontrivial $V(\Phi)$ has alternate sign and (iii) wormholes with two asymptotically flat regions and mirror symmetric with respect to its throat has zero Schwarzschild mass at both asymptotics \cite{bro2}.
Recently, Bronnikov et al. \cite{bro3} proposed an interesting theorem termed as ``{\it no-go theorem}". It states that existence of wormhole solutions with flat and/or AdS asymptotic regions on both sides of the throat is impossible if the matter source is isotropic. \\

Usually four  fundamental  ingredients (three from spacetime metric and one from matter distribution part) are  mandatory to  form  a traversable wormhole. Consider the wormhole structure is characterized by the static, spherical symmetric metric
\begin{equation}
ds^2=e^{2f(r)}dt^2-\left[1-\frac{b(r)}{r}\right]^{-1}dr^2-r^2(d\theta^2+\sin^2\theta ~d\phi^2). \label{metr}
\end{equation}
Here the functions $f(r)$ and $b(r)$ are known as redshift and shape functions  respectively.

Here the four criteria are
\begin{enumerate}

\item[(1)] The redshift function, $f(r)$ should be finite to avoid  an event horizon.  This also indicates that  the wormhole spacetime is asymptotically nonflat.

\item[(2)] The shape function,  $b(r)$,  satisfies the flare-out conditions   at  the  throat  $r  =  r_{th}~ : ~ b(r_{th} ) ~~ =  r_{th}$   and $b'(r_{th} ) < 1$.

\item[(3)]  $ b(r)$ should be less than $r$ for $r > r_{th} $.

\item[(4)]  Violation of  the null energy condition (NEC) near the throat, i.e., $T_{\mu\nu}u^\mu u^\nu    <  0$, where $T_{\mu\nu}$  is the stress-energy tensor and $u^\mu$  is any future directed null vector. This indicates that $\rho + P_r  < 0$  near the throat, where $\rho$ and $P_r$ are energy density and radial pressure respectively.

\end{enumerate}

Recently it was proposed  that wormholes could exist in both the outer regions of the galactic halo and in the central parts of the halo. These studied were  based on two possible choices for the dark matter density profiles (NFW or Navarro-Frenk-White) density profile and the URC (Universal Rotation Curves)  simulation and fittings \cite{16,17,18}. The second  result is an important compliment to the earlier result and  thereby confirming the possible existence of wormholes in most of the spiral galaxies with the universally distributed dark matter. These theoretical results  provide an incentive for scientists to seek observational evidence for wormholes in the galactic halo region. Therefore, it is interesting to obtain  the clear expressions of the physical parameters  of the obtained wormhole spacetime structure and to study their properties and characteristics. Recently, Herrera et al \cite{1} discovered a new algorithm  to obtain all static spherically symmetric anisotropic perfect fluid solutions. As wormholes are exact, anisotropic perfect fluid solutions of Einstein field equations, we adopt the Herrera's approach to find the generating functions comprising the wormhole like geometry.

  Our goal in this work is  to  find the generating functions comprising the wormhole like geometry as well as to  construct wormhole like  geometry from given generating functions.  According to Herrera et al discovery,  there are two generating functions to describe all static spherically symmetric  comprising anisotropic matter distribution.    Using this notion, we will show that  there are two generating functions  needed for  wormhole construction:  one ($Z(r)$) is related to the geometry of spacetime and other ($\Pi(r)$) is related to the matter distribution comprising the wormhole.  In this work, for the   first look , we have used  some specific forms of shape and redshift functions to find an idea about the nature of the generating functions. Note that some interesting properties are found of generating functions as follow:  the generating function  $Z(r)$ related to the redshift  function is always positive and decreasing function of radial coordinate and the second generating function ($\Pi(r)$) is always negative and increasing in nature.  Both functions are asymptotic in nature for large r.   Note that the obtained generating function  ($\Pi(r)$) is related to the matter distribution, therefore, it plays a crucial role to check  the violation of NEC  (which is one of the essential condition for static traversable wormholes). We are also trying to provide a prescription of generating function ($\Pi(r)$, negative and increasing in nature)  to obtain   the shape functions of the wormholes.  By monitoring the measure of anisotropy ($\Pi(r)$) of the matter distribution,  one can get   wormholes. That means one requires the knowledge of measure of anisotropy to generate some wormholes.

The paper is organized as follows. In  Sec. II, we construct the gravitational equations with  general anisotropic energy momentum tensor. In Sec. III, the  generating functions of the wormhole solutions have been found and  consequently try to find generating functions corresponding to different values of redshift $f(r)$ and shape functions  $b(r)$ of the wormholes as well as shape function for given generating functions. Finally some concluding remarks have been made in the next section.

\section{The Einstein field equations}

Let us consider the matter distribution comprising the wormhole is characterized by the  general anisotropic energy momentum tensor
\begin{equation}
T_\nu^\mu = (\rho + P_t)u^{\mu}u_{\nu} - P_t g^{\mu}_{\nu}+(P_r -P_t )\eta^{\mu}\eta_{\nu}, \label{ten}
\end{equation}
with $u^{\mu}u_{\mu} = - \eta^{\mu}\eta_{\mu} = 1$, where $\eta^{\mu}$ is the  the unit spacelike vector in the radial direction,  $P_t$ and $P_r$  are the transverse and radial pressures, respectively.

The Einstein field equations for the above metric (\ref{metr})  and energy momentum tensor (\ref{ten}) are

\begin{eqnarray}
8\pi\rho &=& \frac{b^\prime(r)}{r^2}, \label{dens}\\
8\pi P_r &=& \left(1-\frac{b(r)}{r}\right)\left(\frac{1}{r^2}+\frac{2f^\prime(r)}{r}\right)-\frac{1}{r^2}, \label{pre}\\
8\pi P_t &=& \left(1-\frac{b(r)}{r}\right)f^{\prime\prime}(r)+\left(1-\frac{b(r)}{r}\right){f^\prime}^2(r)+\frac{1}{2}\left(\frac{b(r)}{r^2}-\frac{b^\prime(r)}{r}\right)f^\prime(r) + \frac{1}{r}\left(1-\frac{b(r)}{r}\right)f^\prime(r) \nonumber\\
&& +\frac{1}{2r}\left(\frac{b(r)}{r^2}-\frac{b^\prime(r)}{r}\right). \label{pt}
\end{eqnarray}

Here prime denotes derivative with respect to $r$.

\section{Generating functions}

Recently, Herrera et al \cite{1} developed a formalism to obtain all static spherically symmetric   solutions of locally anisotropic fluids. This new formalism needs the knowledge of two  functions, known as generating functions to generate all possible solutions.

Now, we present the general equations and the formalism to obtain the solutions of the metric potentials in terms of two generating functions.

From equations (\ref{pre})and (\ref{pt}), we obtain
\begin{eqnarray}
8\pi (P_r-P_t) &=& \left(1-\frac{b(r)}{r}\right)\left[-f^{\prime\prime}(r)-{f^\prime}^2(r)+\frac{f^\prime(r)}{r}+\frac{1}{r^2}\right] +\left(\frac{b^\prime(r)}{r}-\frac{b(r)}{r^2}\right)\left(f^\prime(r)+\frac{1}{r}\right)-\frac{1}{r^2}.
\end{eqnarray}

Now, we introduce  the following variables
\begin{equation}
e^{2f(r)}=\exp \left(\int\left[2Z(r)-{2 \over r}\right]dr\right),~~~\mbox{and}~~~1-\frac{b(r)}{r}=y(r).\label{gen1}
\end{equation}

Equation (\ref{gen1}) yields
\begin{equation}
Z(r)= f^\prime(r)+ \frac{1}{r}.
\end{equation}

Let us denote a  function $\Pi(r)$ as
\begin{equation}
\Pi(r)=8\pi(P_r-P_t)=y(r)\left(-Z^\prime(r)-Z^2(r)+\frac{3Z(r)}{r}-\frac{2}{r^2}\right)-y^\prime(r) Z(r)-\frac{1}{r^2}. \label{eom}
\end{equation}
Equation (\ref{eom}) follows
\begin{equation}
y^\prime(r)+y\left(\frac{Z^\prime(r)}{Z(r)}+Z(r)-\frac{3}{r}+\frac{2}{r^2Z(r)}\right)=-\frac{1}{Z(r)}\left(\Pi(r)+\frac{1}{r^2}\right).\label{eo1}
\end{equation}

After solving equation (\ref{eo1}), we obtain for $b(r)$
\begin{eqnarray}
b(r)=r-\frac{r^4}{Z(r)~\exp \left[\int\left(\frac{2}{r^2Z(r)}+Z(r)\right)dr\right]}~\left[C_1-\int \frac{1+\Pi(r) r^2}{r^5}~\exp\left(\int\left\{\frac{2}{r^2Z(r)}+Z(r)\right\}dr\right)~dr\right], \label{th}
\end{eqnarray}
where $C_1$ is a integration Constant.\\

Therefore, using equations (\ref{gen1}) and (\ref{th}) in (\ref{metr}),  we get
\begin{eqnarray}
ds^2 &=& \exp \left[\int\left(2Z(r)-\frac{2}{r}\right)dr\right]~dt^2 - \frac{Z(r)~\exp \left[\int\left\{\frac{2}{r^2Z(r)}+Z(r)\right\}dr \right]dr^2}{r^3\left[C_1-\int\frac{1+\Pi(r) r^2}{r^5} ~\exp \left(\int\left\{\frac{2}{r^2 Z(r)}+Z(r)\right\}dr\right)dr\right]}  \nonumber\\
&& -r^2(d\theta^2+sin^2\theta d\phi^2).
\end{eqnarray}
Hence one can obtain any solution describing a static anisotropic fluid distribution   by means of the two generating functions $\Pi(r)$  and $Z(r)$.

The expression for NEC can be expressed in terms of the two generating functions $Z(r)$ and $\Pi(r)$ as
\begin{eqnarray}
8\pi(\rho+P_r) &=& \frac{1}{r \alpha (r) Z(r)^2} \Bigg[r Z(r)^2 \left\{C_1 r^4+\alpha (r)\right\}+Z(r) \left\{-4 C_1 r^4+r^2 \alpha (r) \Pi (r)+\alpha (r)\right\} \nonumber \\
&& +C_1 r^3 \left\{r^2 Z'(r)+2\right\}-r^3 \left\{r^2 Z'(r)+r^2 Z(r)^2-4 r Z(r)+2\right\} \times \nonumber \\
&& \int \frac{\alpha (r) \left\{r^2 \Pi (r)+1\right\}}{r^5}~ dr \Bigg]
\end{eqnarray}
where $\alpha(r) = \exp \left[\int \left(\frac{2}{r^2 Z(r)}+Z(r)\right) ~ dr\right]$.

In the following text, we try to find out the generating functions corresponding to different values of redshift ($f(r)$) and shape functions  ($b(r)$) of the wormholes as well as shape function for given generating functions.

\subsection{Generating function corresponding to $f(r)=0$: }

Now, we will consider three cases with different shape functions for  constant redshift function, so called tidal force solution. This  choice is acceptable as the redshift function $f(r)$ is always finite for all values of $r$ to evade an event horizon.

\subsubsection{Shape function: $b(r)=r_{0}~(\frac{r}{r_{0}})^{n}$ .}

For this type of  shape function, $b(r)=r_{0}~(\frac{r}{r_{0}})^{n}$, we have $r_0$ is the throat radius. Here  $n$ is an arbitrary constant with the restriction  $n<1$ to satisfy the flaring out condition.

We have found the generating functions as

\begin{equation}
Z(r)=\frac{1}{r} ~~~\mbox{and}~~~\Pi(r)=(n-2)r_o^{1-n}r^{n-3}.
\end{equation}
Since the violation of NEC is one of the essential condition for static traversable wormholes, we calculate $ 8\pi(P_r+\rho)$ as
\begin{equation}
8\pi(P_r+\rho)=(n-1)r_o^{1-n}r^{n-3}.
\end{equation}
Here, the violation of NEC is clear as the parameter $n$ is less than 1, the above equation gives $P_r+\rho<0$. We also draw the plots given in figure 1. The figure indicates that the generating function $Z(r)$ related to the geometry of the spacetime is always positive and converges for large $r$. However, generating function $\Pi(r)$ related to the matter distribution is always negative. Here the NEC is violated as expected (see figure 1).

\begin{figure*}[thbp]
\begin{center}
\begin{tabular}{rl}
\includegraphics[width=5.cm]{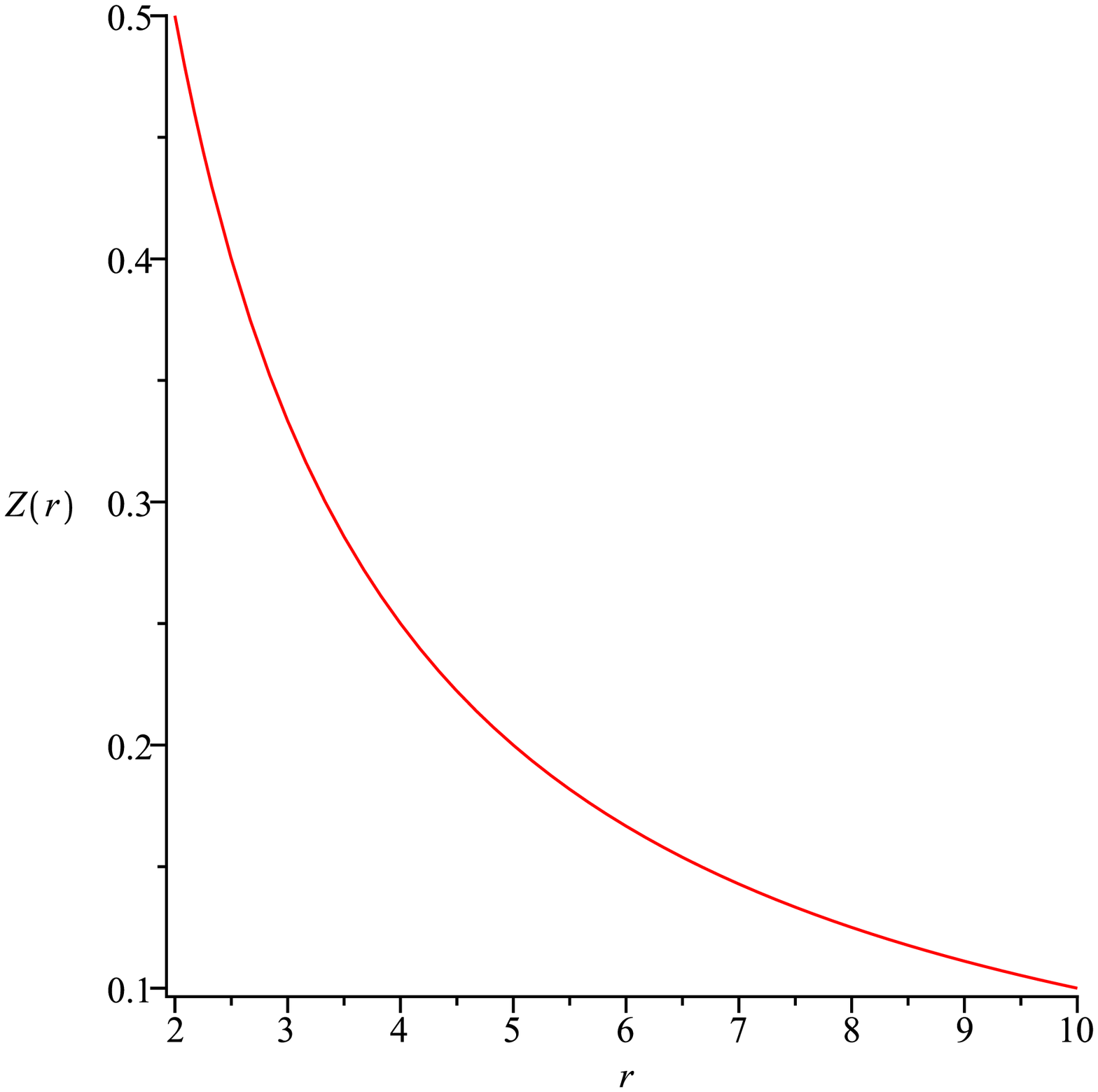}&
\includegraphics[width=5.cm]{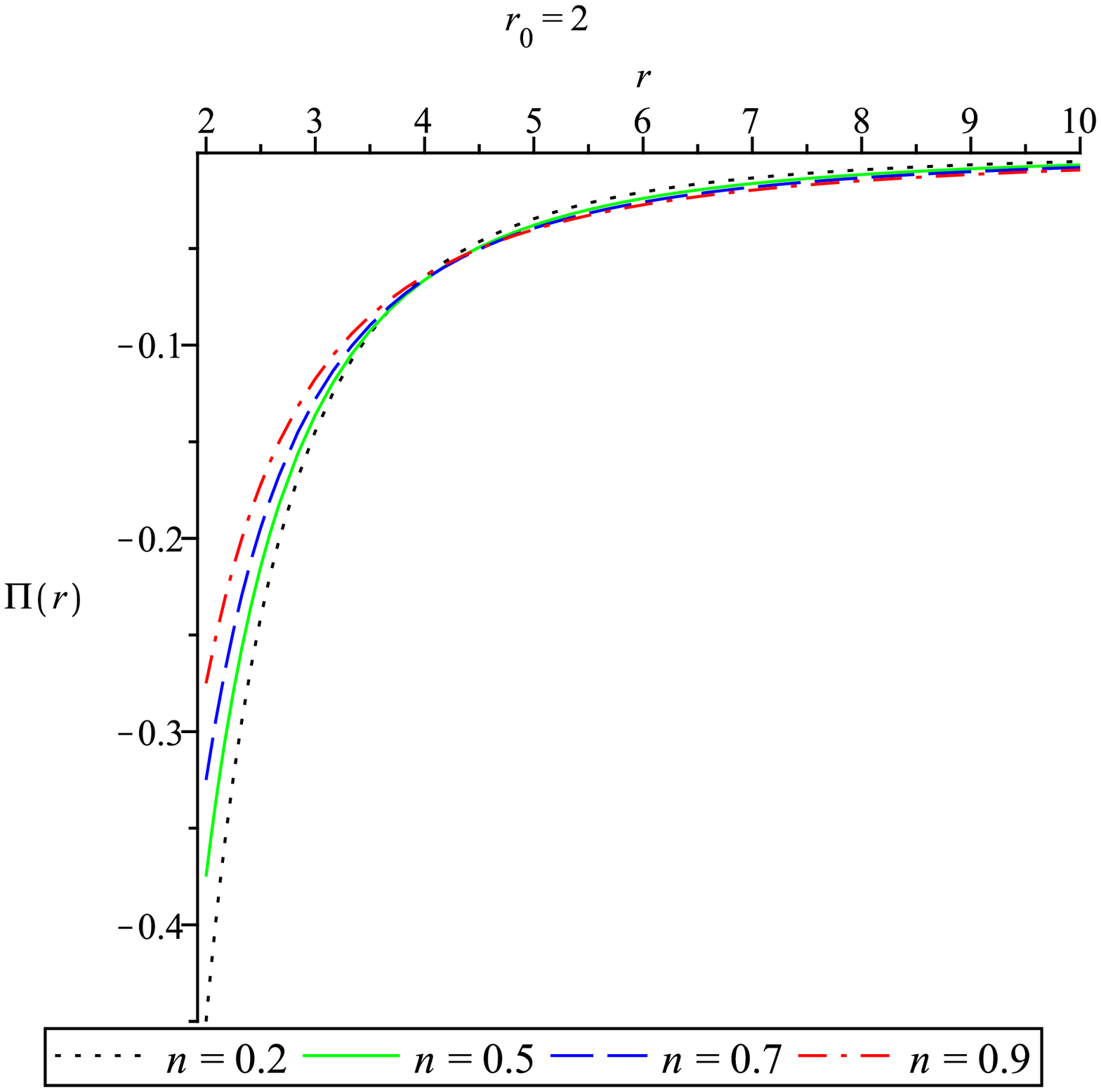}
\includegraphics[width=5.cm]{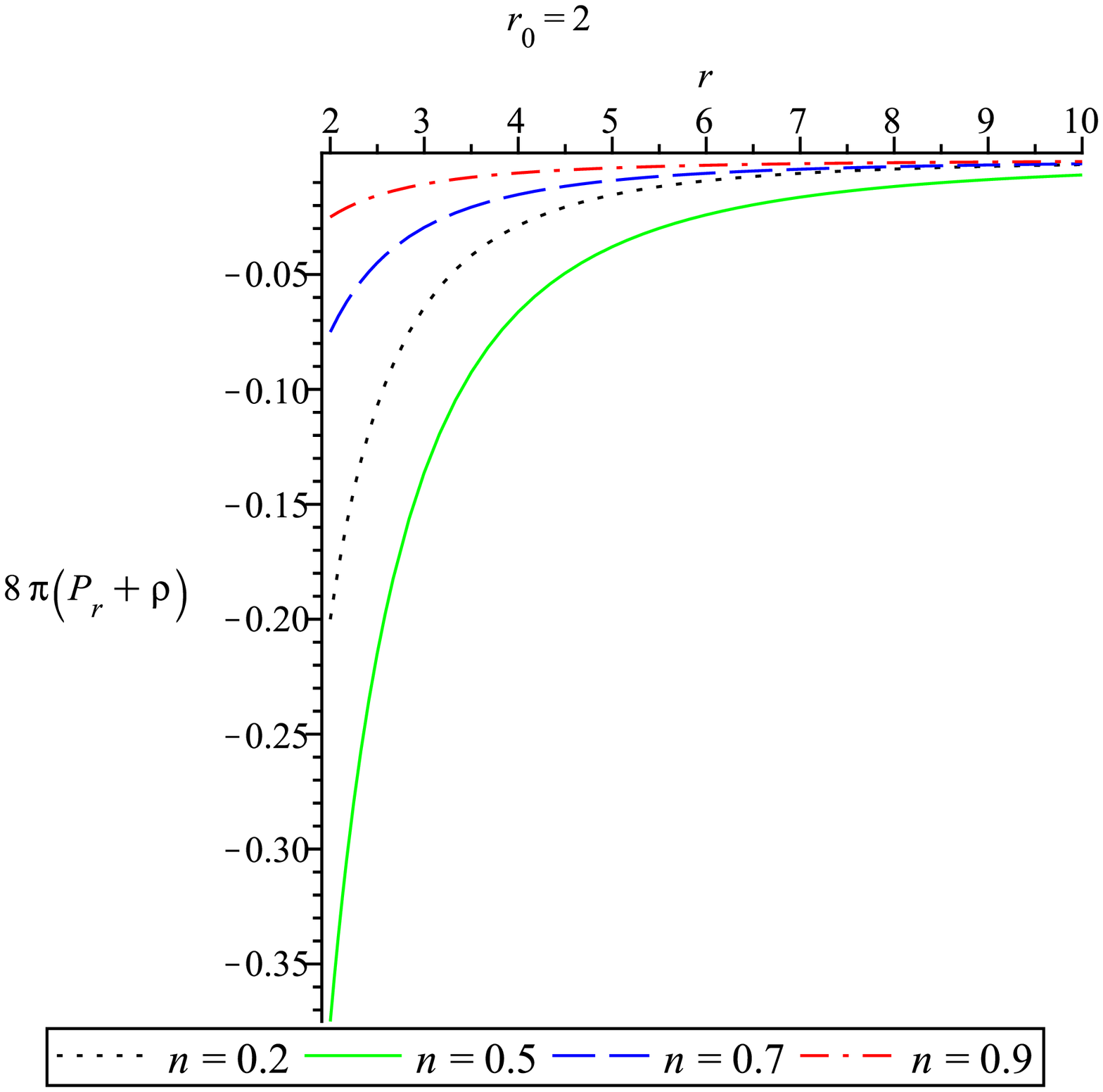}
\\
\end{tabular}
\end{center}
\caption{(left) Generating function $Z$ is always positive.    (middle) Generating function $\Pi$ is always negative.   (right) NEC  is violated  .}
\end{figure*}

\subsubsection{Shape function:  $b(r)=A~ \tan^{-1}(Cr)$.}

Here, we consider the  shape function  $b(r)=A ~\tan^{-1}(Cr)$, where $A$ and $C$ are constants. This choice has  one important advantage over the previous case for polynomial function of $r$ since $b(r)/r\rightarrow 0$ more rapidly. However, we have checked graphically (see figure 2) whether this shape function satisfies all criteria of wormhole. Note that here, we have throat as well flare-out condition is satisfied.

As previous case, generating function $\Pi(r)$   related to the matter distribution is always negative and  the NEC is violated (see figure 3).
 \begin{figure*}[thbp]
\begin{center}
\begin{tabular}{rl}
\includegraphics[width=5.cm]{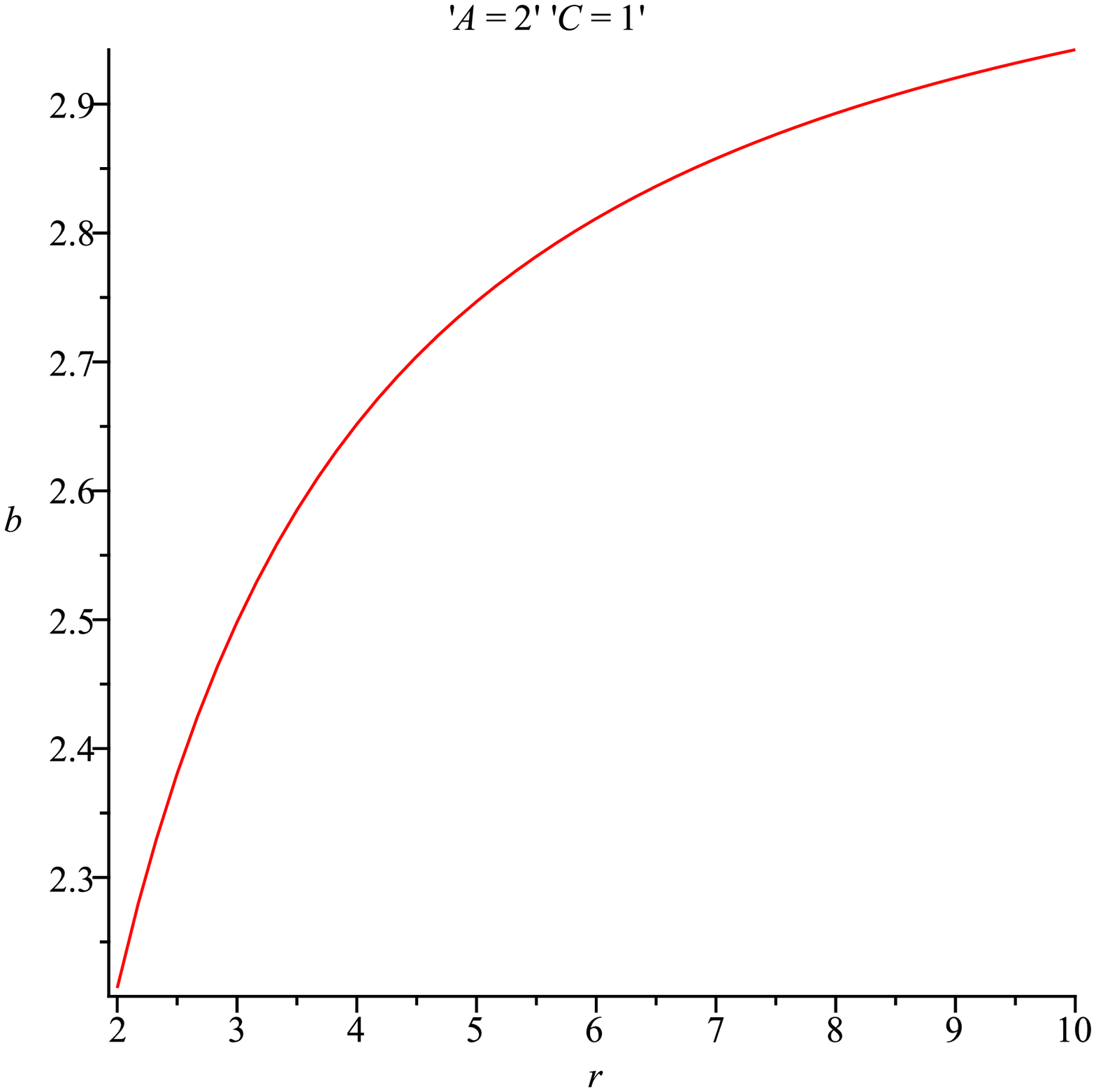}&
\includegraphics[width=5.cm]{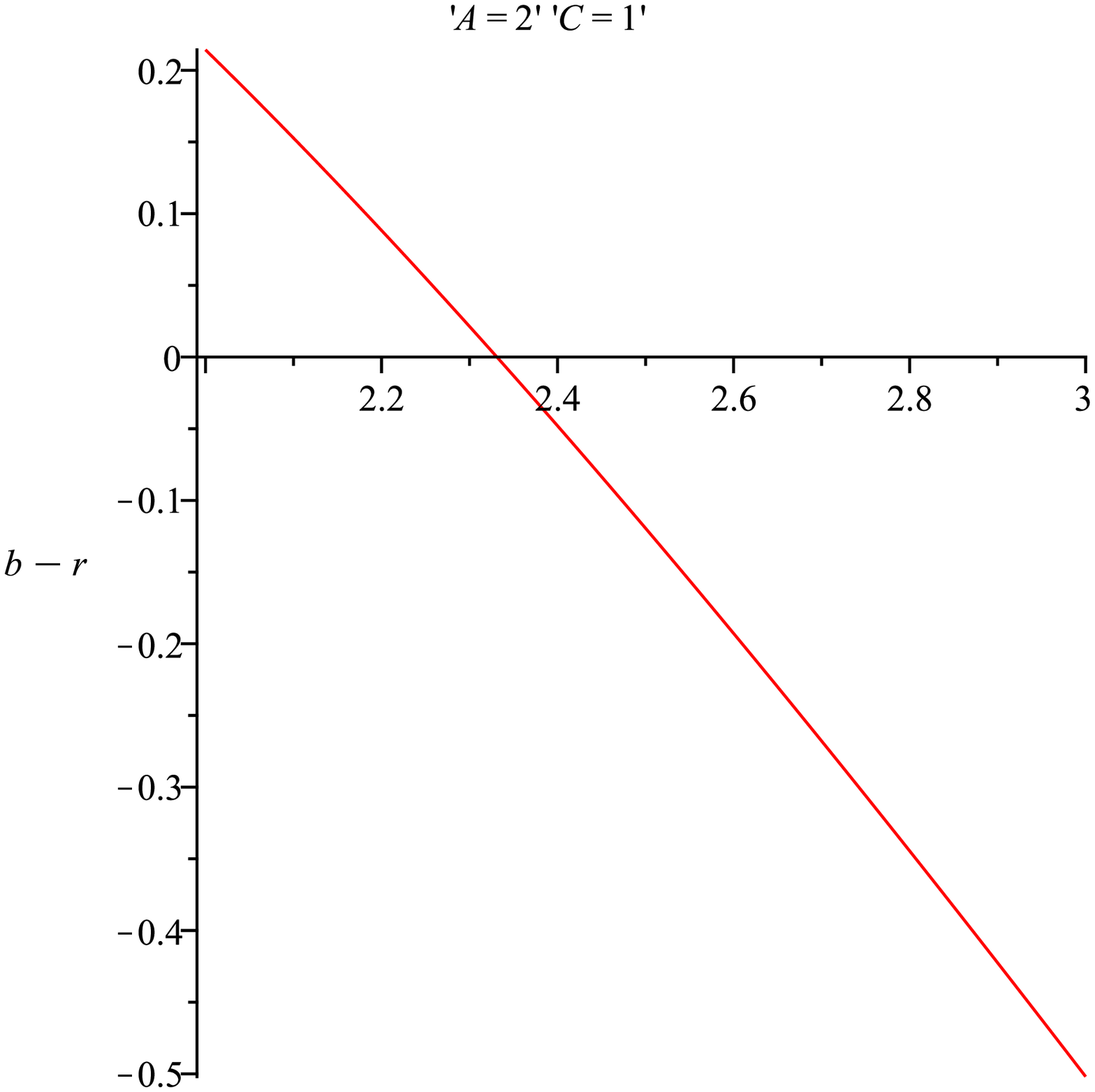}
\includegraphics[width=5.cm]{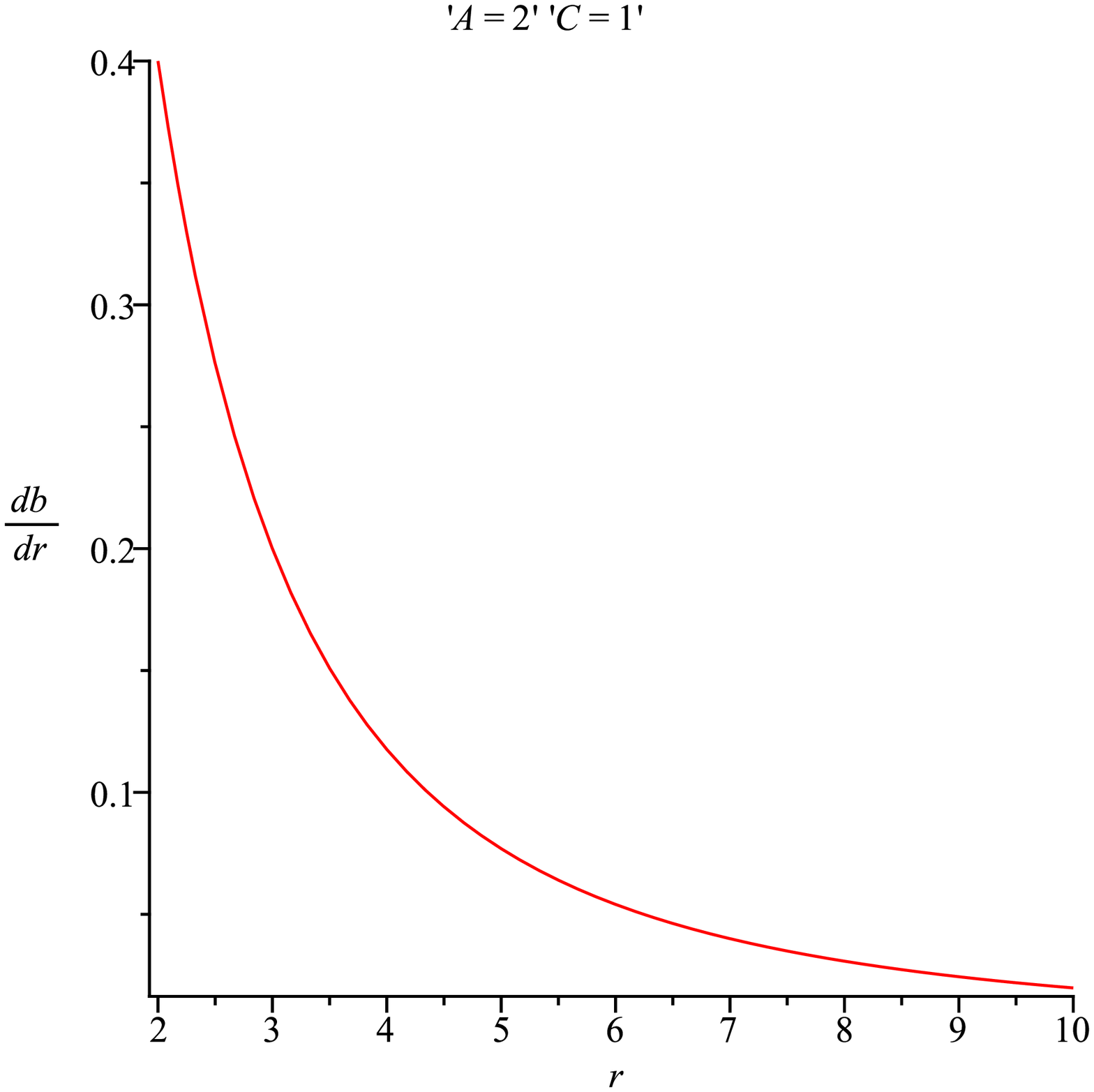}
\\
\end{tabular}
\end{center}
\caption{(left) Shape function of  wormhole.    (middle) Throat of  wormhole is located where the $b(r) -r $ cuts $r$ axis.  (right) Flare-out condition  is satisfied .}
\end{figure*}

The generating functions are
\begin{equation}
Z(r)=\frac{1}{r} ~~~\mbox{and}~~~\Pi(r)=\frac{AC}{r^2(1+C^2r^2)}-\frac{2A \tan^{-1}(Cr)}{r^3}.
\end{equation}
Also we find
\begin{equation}
8\pi(P_r+\rho)=\frac{A}{r^2}\left(\frac{C}{1+C^2r^2}-\frac{\tan^{-1}(Cr)}{r}\right).
\end{equation}

\begin{figure*}[thbp]
\begin{center}
\begin{tabular}{rl}
\includegraphics[width=5.cm]{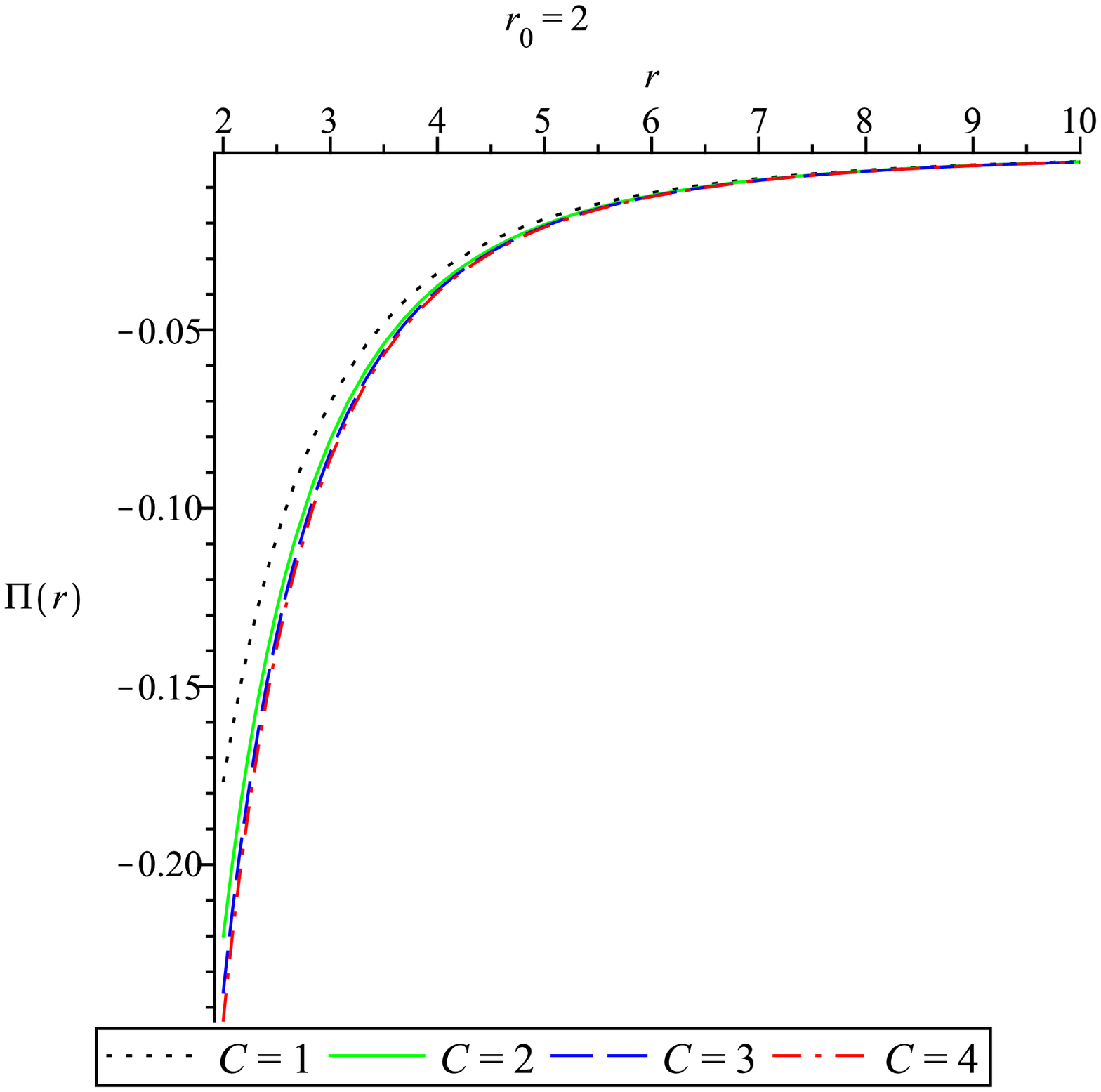}&
\includegraphics[width=5.cm]{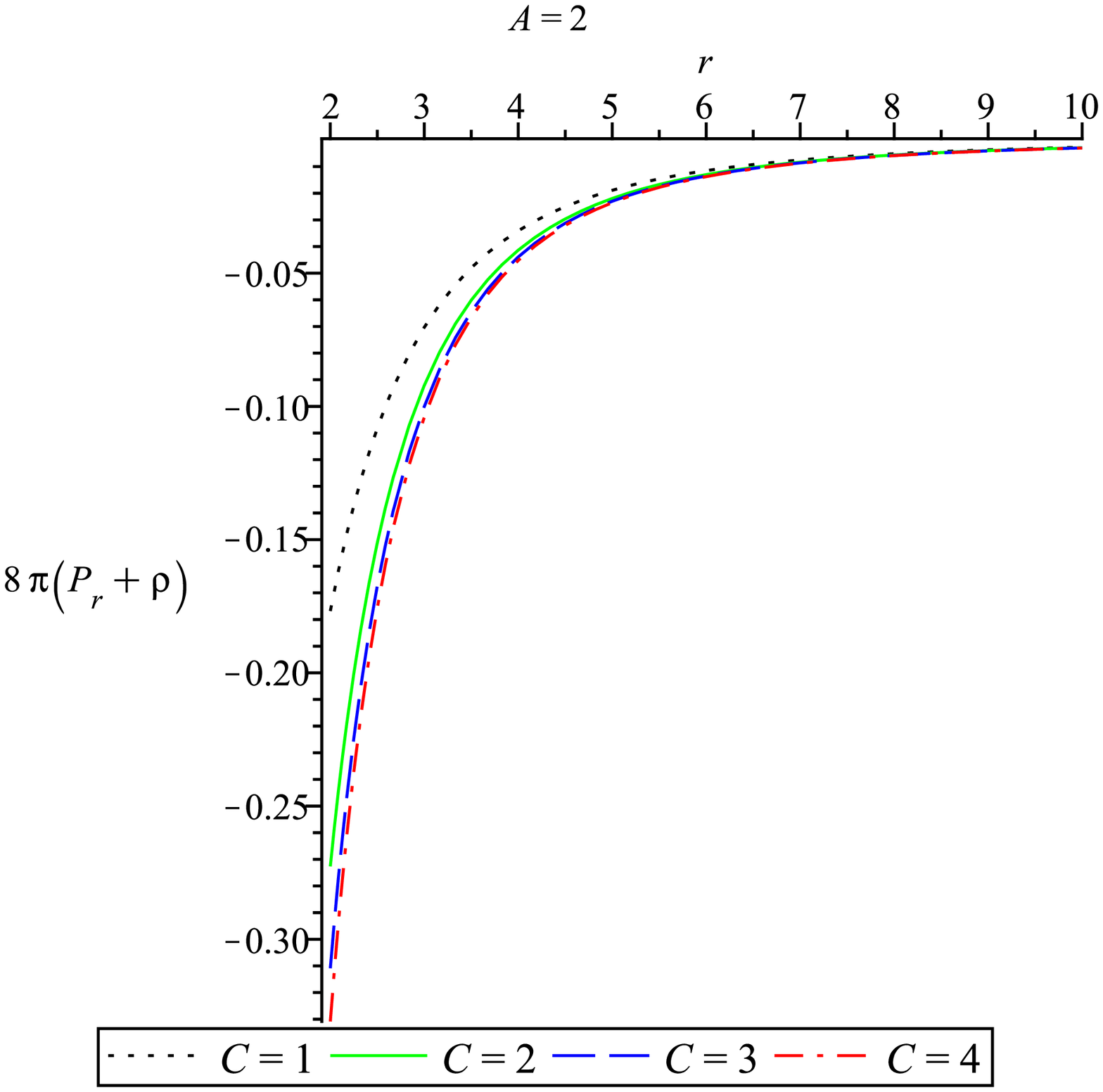}
\\
\end{tabular}
\end{center}
\caption{(left)  Generating function $\Pi$ is always negative.   (right) NEC  is violated  .}
\end{figure*}

\subsubsection{Shape function : $b(r)=r_{0}~[1+\beta^2(1-\frac{r_{0}}{r})]$. }

The  form of the shape function  $b(r)=r_{0}[1+\beta^2(1-\frac{r_{0}}{r})]$, where  $r_0$ is the throat radius, provides a wormhole solution provided the  arbitrary constant $\beta$ has the  restriction $\beta^2${$<$}1 since  $b'(r_0)<1$.

Here the generating functions are
\begin{equation}
Z(r)=\frac{1}{r}~~~\mbox{and}~~~
\Pi (r)=\frac{3\beta^2r_0^2}{r^4}-\frac{2r_{0}(1+\beta^2)}{r^3}.
\end{equation}
The form  $ 8\pi(P_r+\rho)$ is given by
\begin{equation}
8\pi(P_r+\rho)=\frac{2\beta^2r_0^2}{r^4}-\frac{r_0}{r^3}-\frac{\beta^2r_0}{r^3}.
\end{equation}
Note that as before we get the same nature of the generating functions and NEC is violated ( see figure 4 ).

\begin{figure*}[thbp]
\begin{center}
\begin{tabular}{rl}
\includegraphics[width=5.cm]{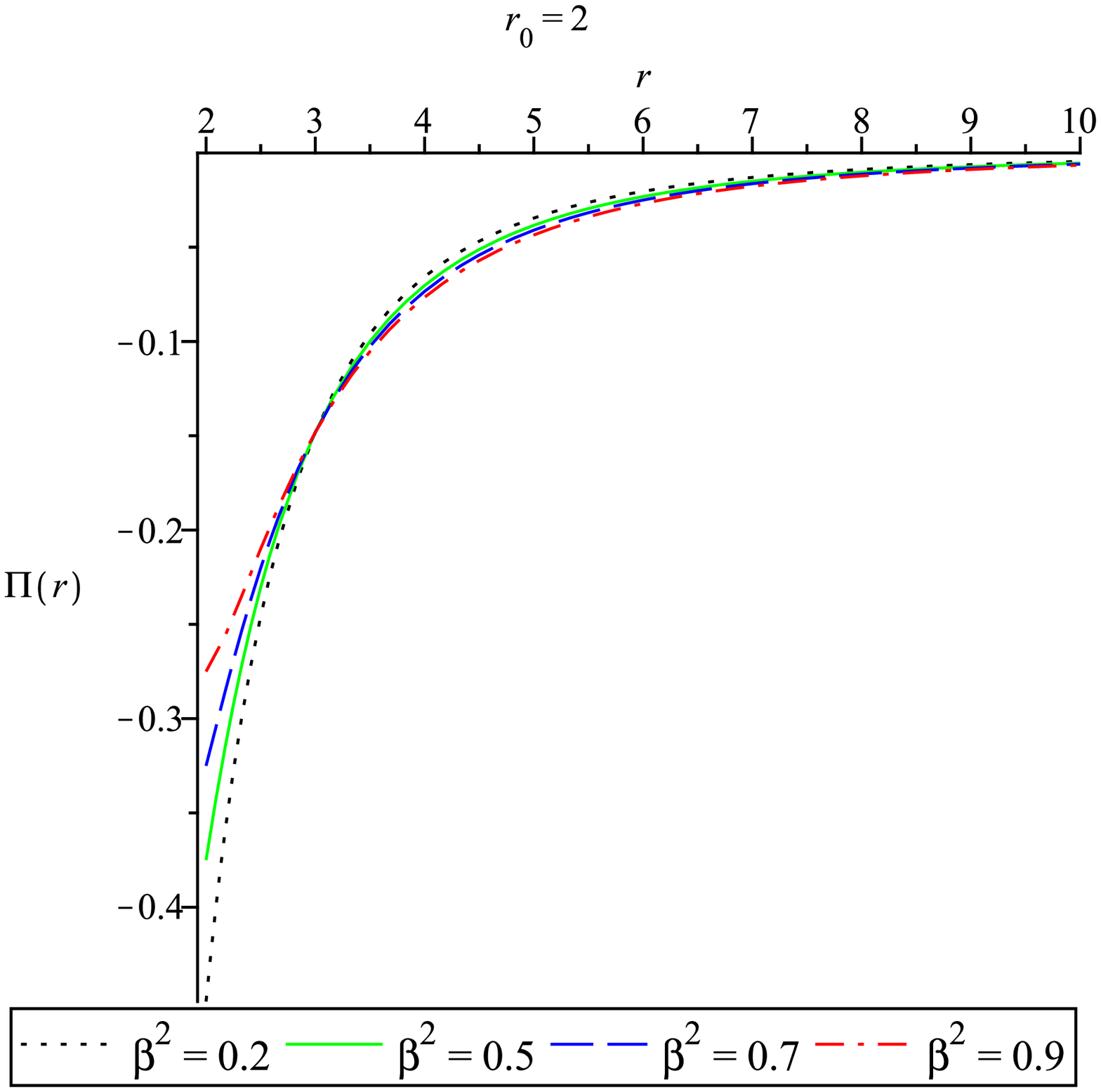}&
\includegraphics[width=5.cm]{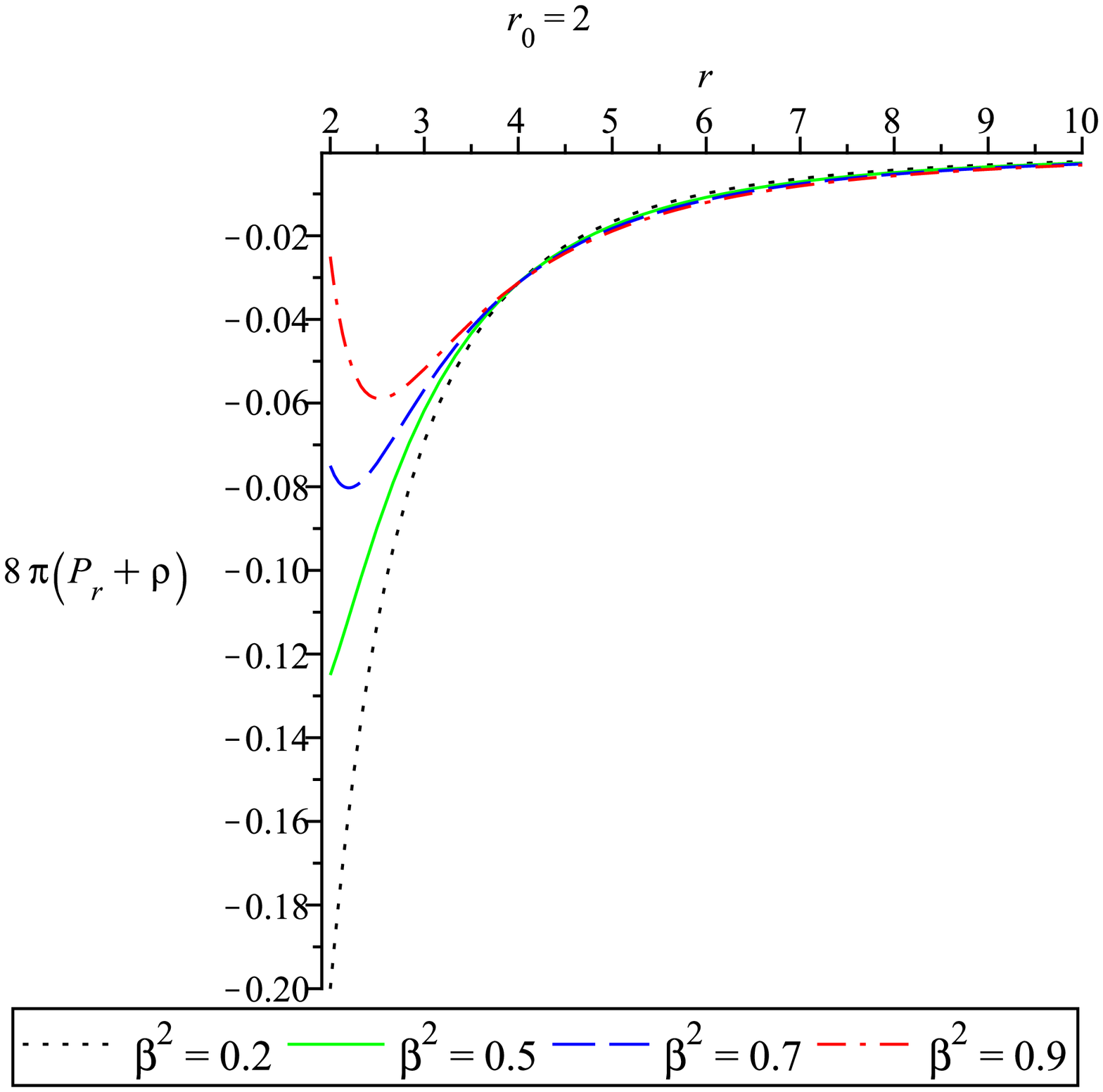}
\\
\end{tabular}
\end{center}
\caption{(left)  Generating function $\Pi$ is always negative.   (right) NEC  is violated .}
\end{figure*}

\subsection{Generating function corresponding to $f(r)=\frac{\alpha}{r}$ :}
 \begin{figure*}[thbp]
\begin{center}
\begin{tabular}{rl}
\includegraphics[width=5.cm]{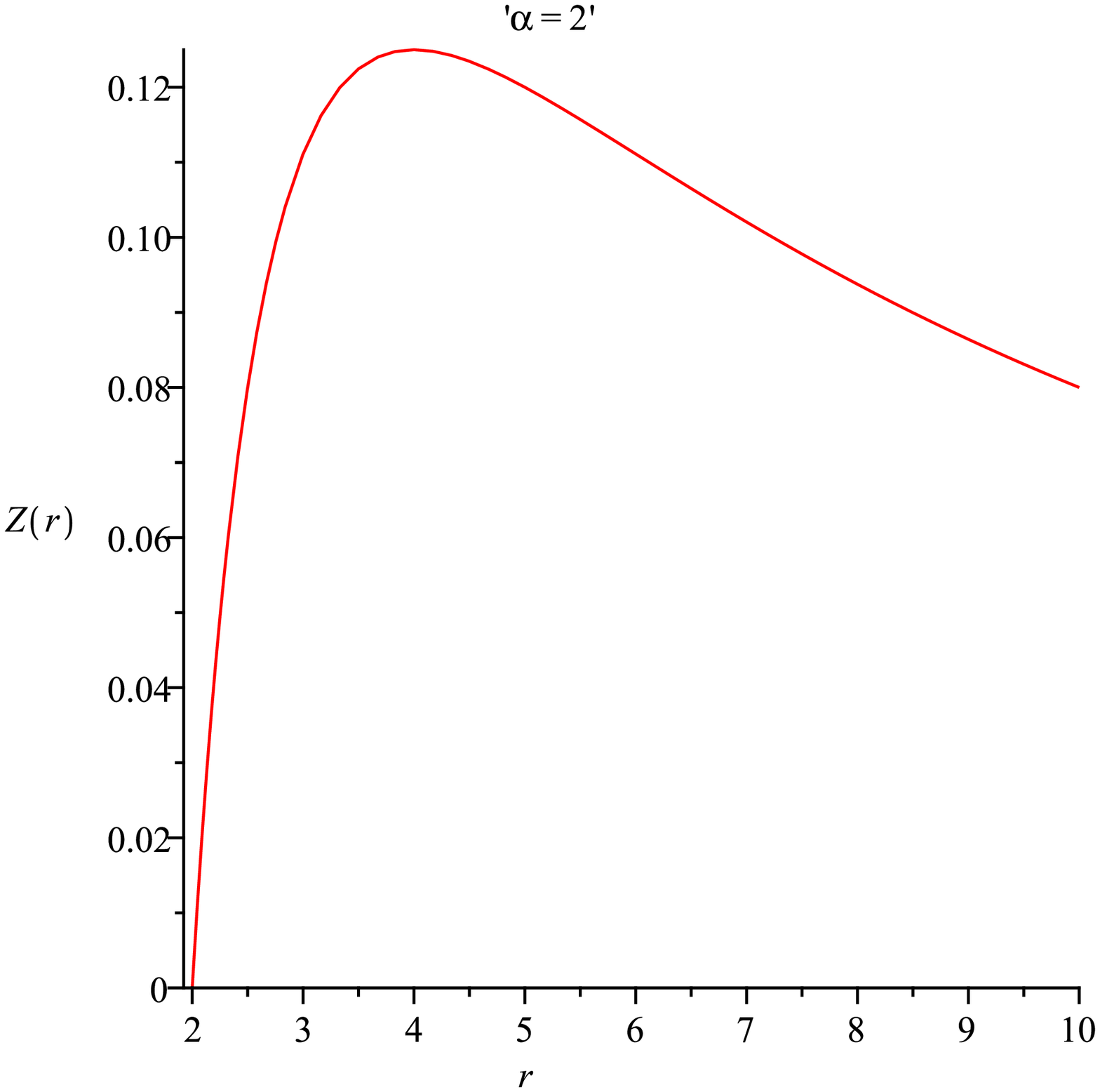}&
\includegraphics[width=5.cm]{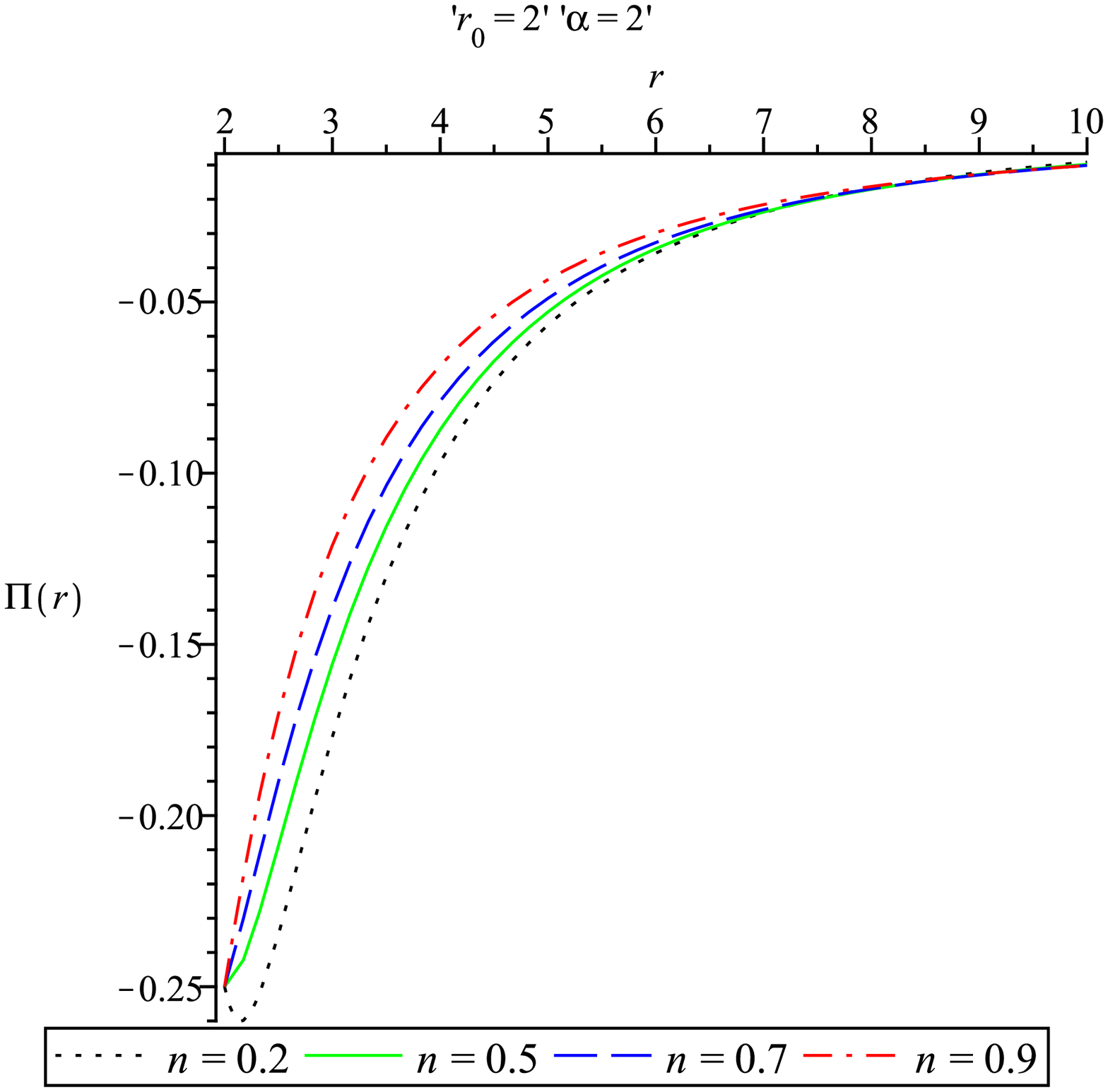}
\includegraphics[width=5.cm]{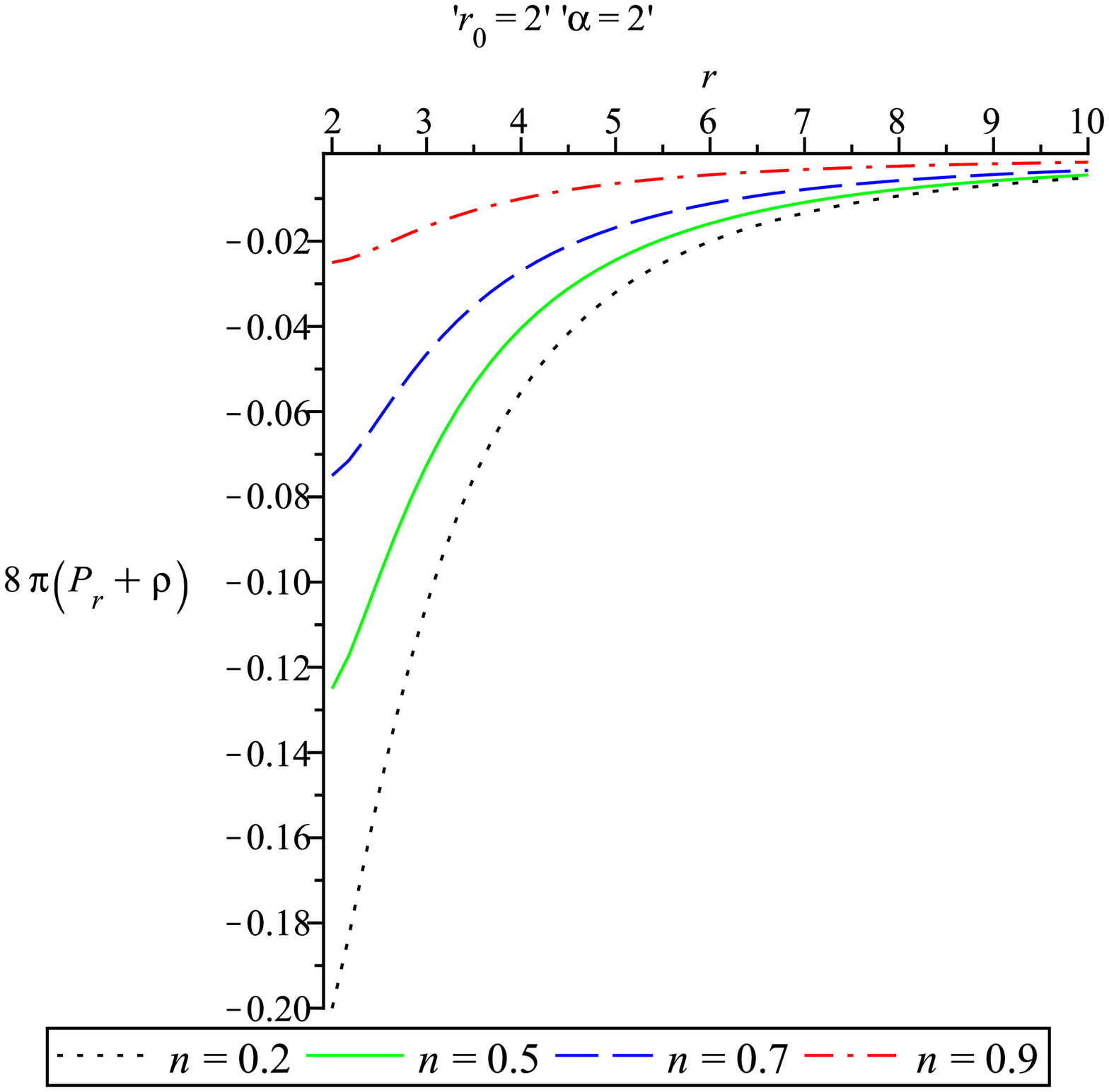}
\\
\end{tabular}
\end{center}
\caption{(left)Generating function $Z$ is always positive.    (middle) Generating function $\Pi$ is always negative.   (right) NEC  is violated .}
\end{figure*}

 \begin{figure*}[thbp]
\begin{center}
\begin{tabular}{rl}
\includegraphics[width=5.cm]{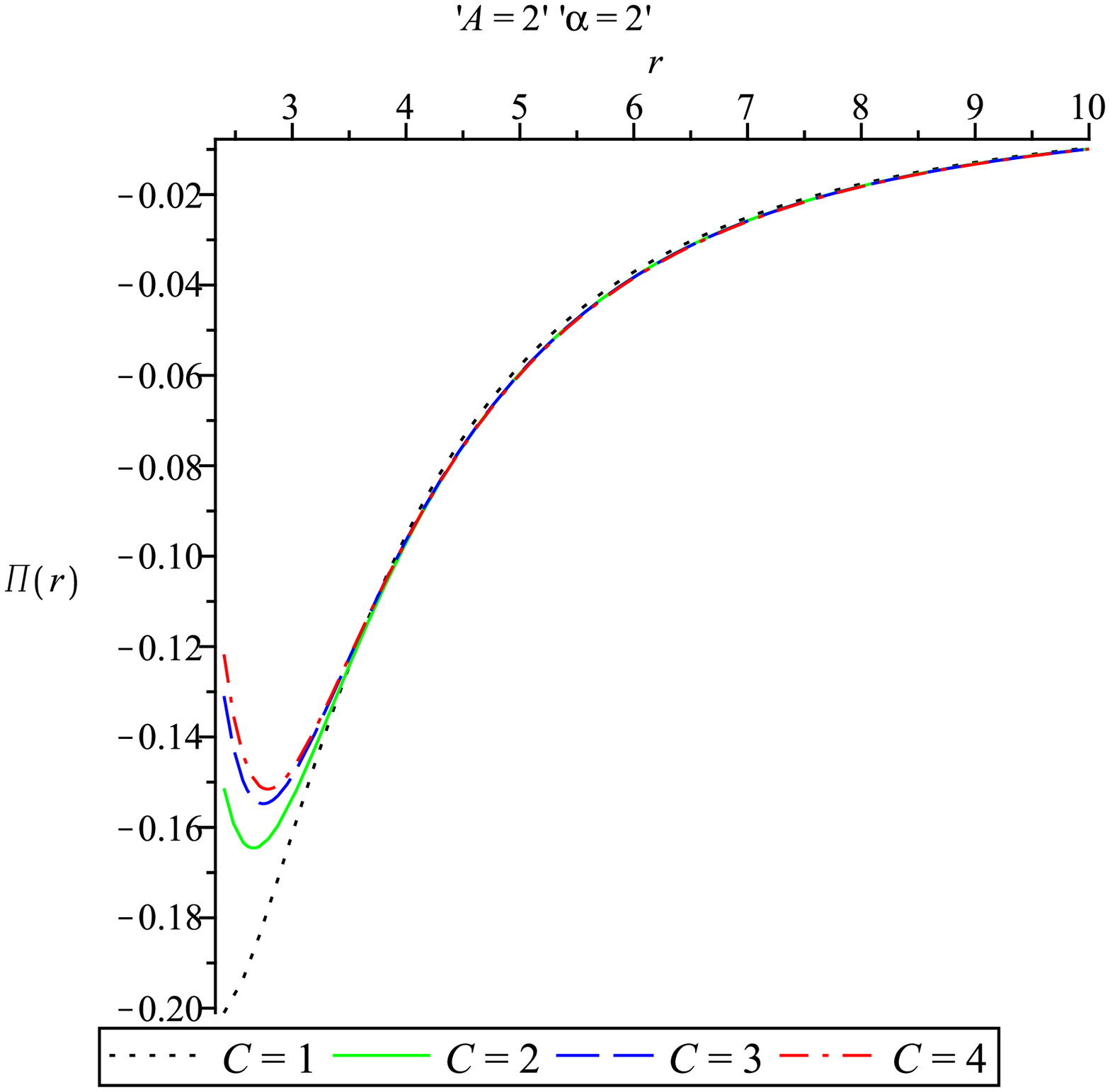}&
\includegraphics[width=5.cm]{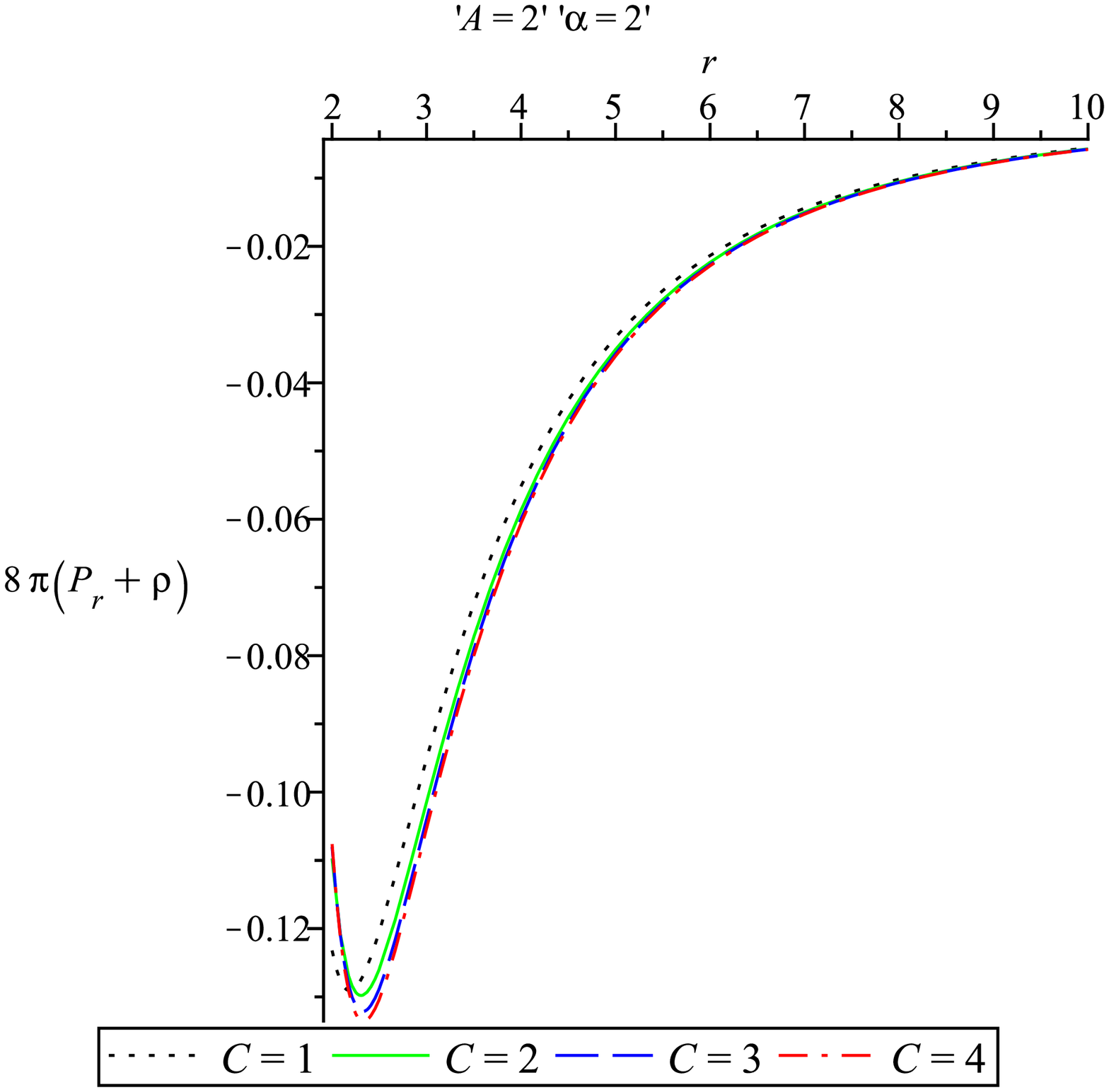}
\\
\end{tabular}
\end{center}
\caption{(left)  Generating function $\Pi$ is always negative.   (right) NEC  is violated .}
\end{figure*}

 \begin{figure*}[thbp]
\begin{center}
\begin{tabular}{rl}
\includegraphics[width=5.cm]{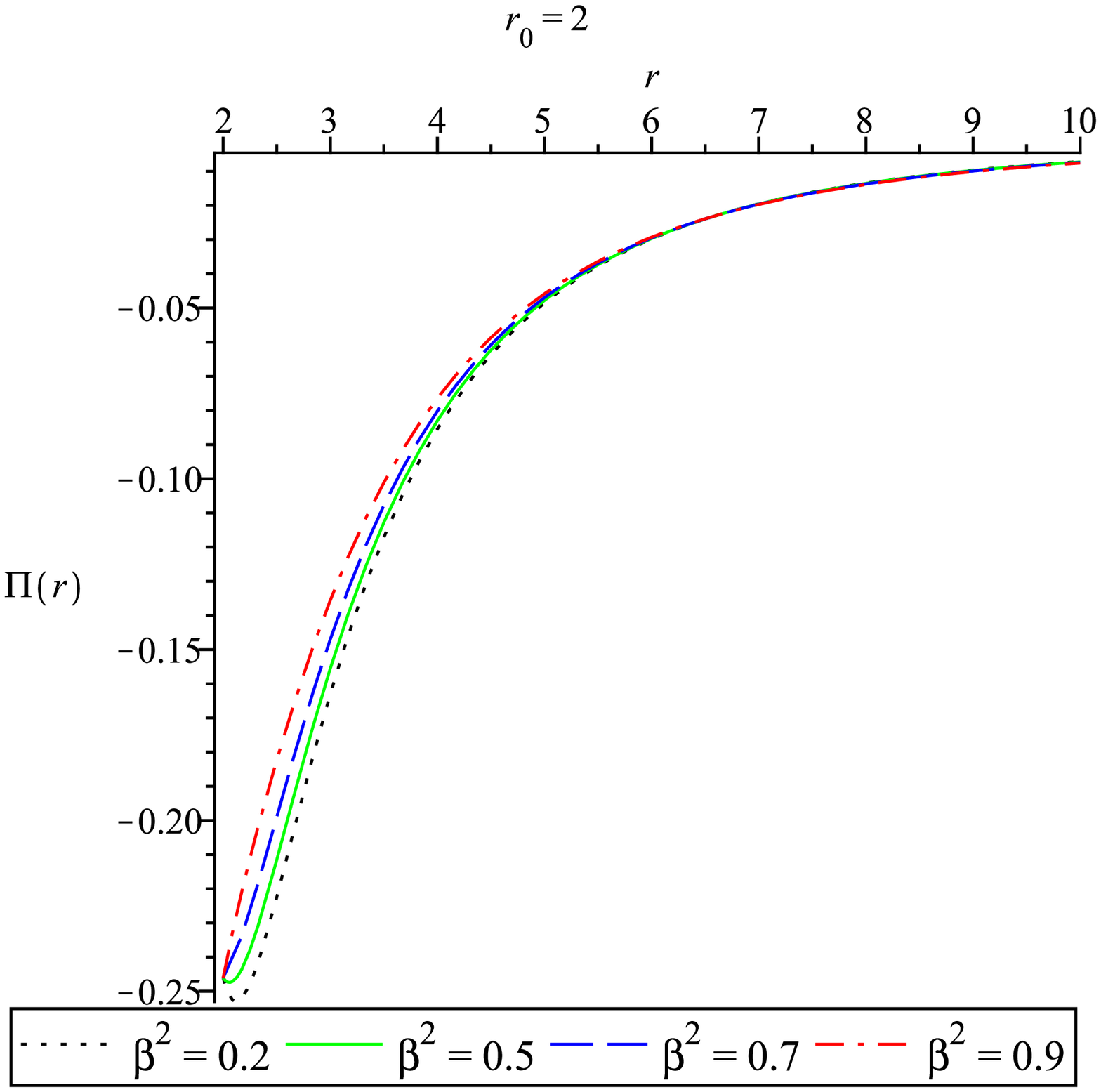}&
\includegraphics[width=5.cm]{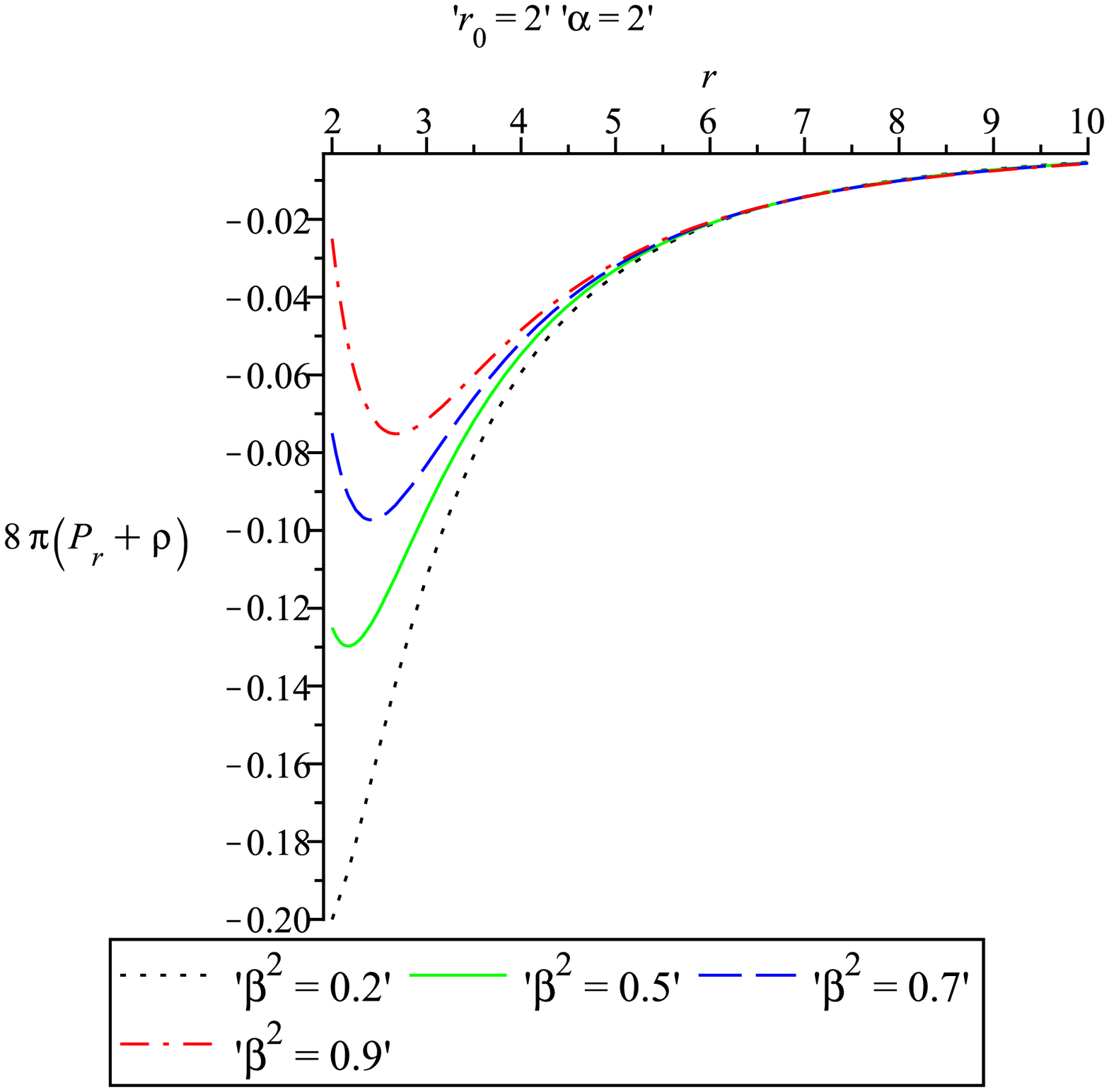}
\\
\end{tabular}
\end{center}
\caption{(left)  Generating function $\Pi$ is always negative.   (right) NEC  is violated .}
\end{figure*}

In this section, we consider the redshift function $f(r)=\frac{\alpha}{r}$, where $\alpha$ is a constant. Note that this redshift function obeys the characteristics of wormhole. Again, we consider the above three forms of shape functions to determine the generating functions separately.  The generating functions and the expression of $8\pi(P_r+\rho)$ for the above three shape functions   are respectively
 \begin{eqnarray}
Z(r) &=& \frac{1}{r}-\frac{\alpha}{r^2}, \\
\Pi (r) &=& -\frac{\alpha}{r^4}-\frac{3\alpha}{r^3}+\alpha^2r_0^{1-n}r^{n-5}+\alpha(4-n)r_0^{1-n}r^{n-4}+(n-2)r_0^{1-n}r^{n-3}, \\
8\pi(P_r+\rho) &=& (n-1)r_o^{1-n}r^{n-3}-\frac{2\alpha}{r^3}+2\alpha r_0^{1-n}r^{n-4}.
\end{eqnarray}

\begin{eqnarray}
\Pi(r) &=& -\frac{\alpha}{r^4}-\frac{3\alpha}{r^3}+\frac{A\alpha^2}{r^5}\tan^{-1}(Cr)+\frac{4A\alpha}{r^4}\tan^{-1}(Cr)-\frac{2A}{r^3}\tan^{-1}(Cr)-\frac{AC}{(1+C^2r^2)}\frac{\alpha}{r^3} \nonumber\\
&& +\frac{AC}{r^2(1+C^2r^2)}, \\
8\pi(P_r+\rho) &=& \frac{AC}{r^2(1+C^2r^2)}-\frac{2\alpha}{r^3}+\frac{A\tan^{-1}(Cr)}{r^3}\left(\frac{2\alpha}{r}-1\right).
\end{eqnarray}

\begin{eqnarray}
\Pi(r) &=& \frac{1}{r^6} \Bigg[\frac{1}{r^2}-\alpha^2\beta^2r_0^2+\alpha r_0 r(\alpha+\alpha \beta^2-3\beta^2 r_0)+r^2\{\beta^2 r_0^2+3\alpha r_0(1+\beta^2)-\alpha^2\} \nonumber \\
&& -r^3\{3\alpha+r_0(1+\beta^2)\} \Bigg],\\
8\pi(P_r+\rho) &=& -\frac{2\alpha \beta^2 r_0}{r^5}+\frac{1}{r^4} \Big(\beta^2 r_0^2+\beta^2 r_0+2\alpha r_0+2\alpha \beta^2 r_0 \Big)-\frac{1}{r^3} \Big(r_0+\beta^2 r_0+2\alpha \Big).
\end{eqnarray}

We observe  that as before we get the same nature of the generating functions and NEC is violated ( see figures 5 - 7 ).

\subsection{Generating function corresponding to $f(r) = \ln \left(\frac{\sqrt{\gamma^2+r^2}}{r}\right)$: }
In this section, we consider another possible redshift function $f(r) = \ln \left(\frac{\sqrt{\gamma^2+r^2}}{r}\right)$ , where $\gamma$ is a constant. Note that this redshift function follows the features of wormhole. As earlier, we consider the above three forms of shape functions to determine the generating functions separately.  The generating functions and the expression of $8\pi(P_r+\rho)$ for the above three shape functions   take the following forms  respectively

\begin{eqnarray}
Z(r) &=& \frac{r}{r^2+\gamma^2}, \\
\Pi (r) &=& \left(1-r_0^{1-n}r^{n-1}\right)\left[\frac{\gamma^2(2r-3)}{r^2(r^2+\gamma^2)}+\frac{1}{r^2}-\frac{3\gamma^2r^2+2\gamma^4}{r^2(r^2+\gamma^2)^2}\right]-\frac{1}{r^2}+\frac{(n-1)r_0^{1-n}r^{n-1}}{r^2+\gamma^2},\\
8\pi(P_r+\rho) &=& nr_o^{1-n}r^{n-3}-\frac{1}{r^2}+\Big(1-r_o^{1-n}r^{n-1}\Big) \left(\frac{1}{r^2}-\frac{2\gamma^2}{r^2(r^2+\gamma^2)} \right).
\end{eqnarray}

\begin{eqnarray}
\Pi(r) &=& \left(1-\frac{A\tan^{-1}(Cr)}{r}\right)\left[\frac{\gamma^2(2r-3)}{r^2(r^2+\gamma^2)}+\frac{1}{r^2}-\frac{3\gamma^2r^2+2\gamma^4}{r^2(r^2+\gamma^2)^2}\right]-\frac{1}{r^2} - A\left(\frac{\tan^{-1}(Cr)}{r}-\frac{C}{1+C^2r^2}\right)\nonumber\\
&& \left(\frac{1}{r^2+\gamma^2}\right),\\
8\pi(P_r+\rho) &=& \frac{AC}{r^2(1+C^2r^2)}-\frac{2\gamma^2}{r^2(r^2+\gamma^2)}-\frac{A\tan^{-1}(Cr)}{r^3}\left(1-\frac{2\gamma^2}{r^2+\gamma^2}\right).
\end{eqnarray}

\begin{eqnarray}
\Pi(r) &=& \left(1-\frac{r_0(1+\beta^2)}{r}+\frac{\beta^2r_0^2}{r^2}\right)\left[\frac{\gamma^2(2r-3)}{r^2(r^2+\gamma^2)}+\frac{1}{r^2}-\frac{3\gamma^2r^2+2\gamma^4}{r^2(r^2+\gamma^2)^2}\right]-\frac{1}{r^2} -\frac{r_0r(1+\beta^2)-2\beta^2r_0^2}{r^2}\nonumber\\
&& \left(\frac{1}{r^2+\gamma^2}\right),\\
8\pi(P_r+\rho) &=& \frac{\beta^2r_0^2}{r^4}-\frac{2\gamma^2}{r^2(r^2+\gamma^2)}-\frac{r_0+\beta^2r_0-\beta^2r_0^2/r}{r^3}\left(1-\frac{2\gamma^2}{r^2+\gamma^2}\right).
\end{eqnarray}

 \begin{figure*}[thbp]
\begin{center}
\begin{tabular}{rl}
\includegraphics[width=5.cm]{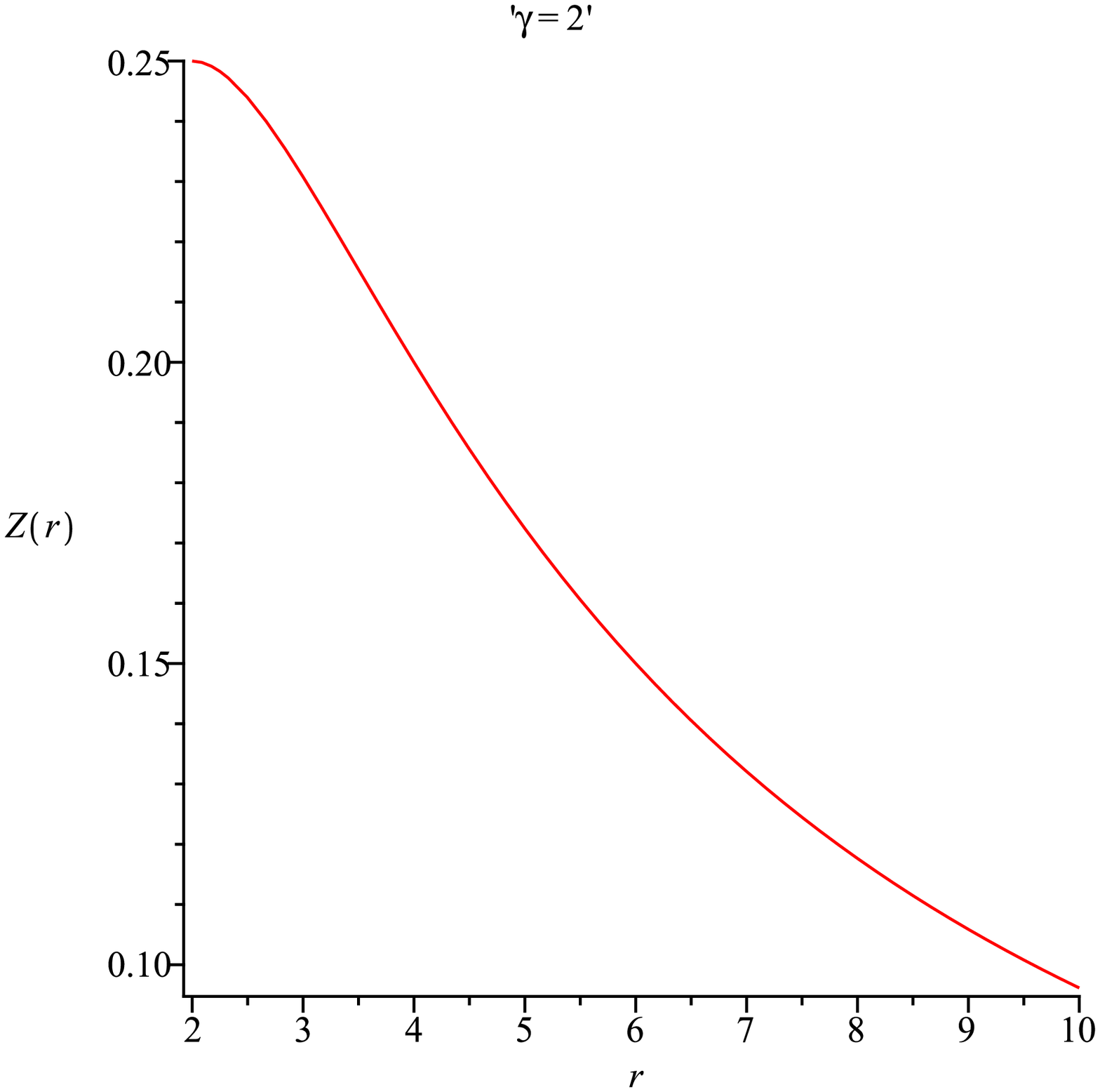}&
\includegraphics[width=5.cm]{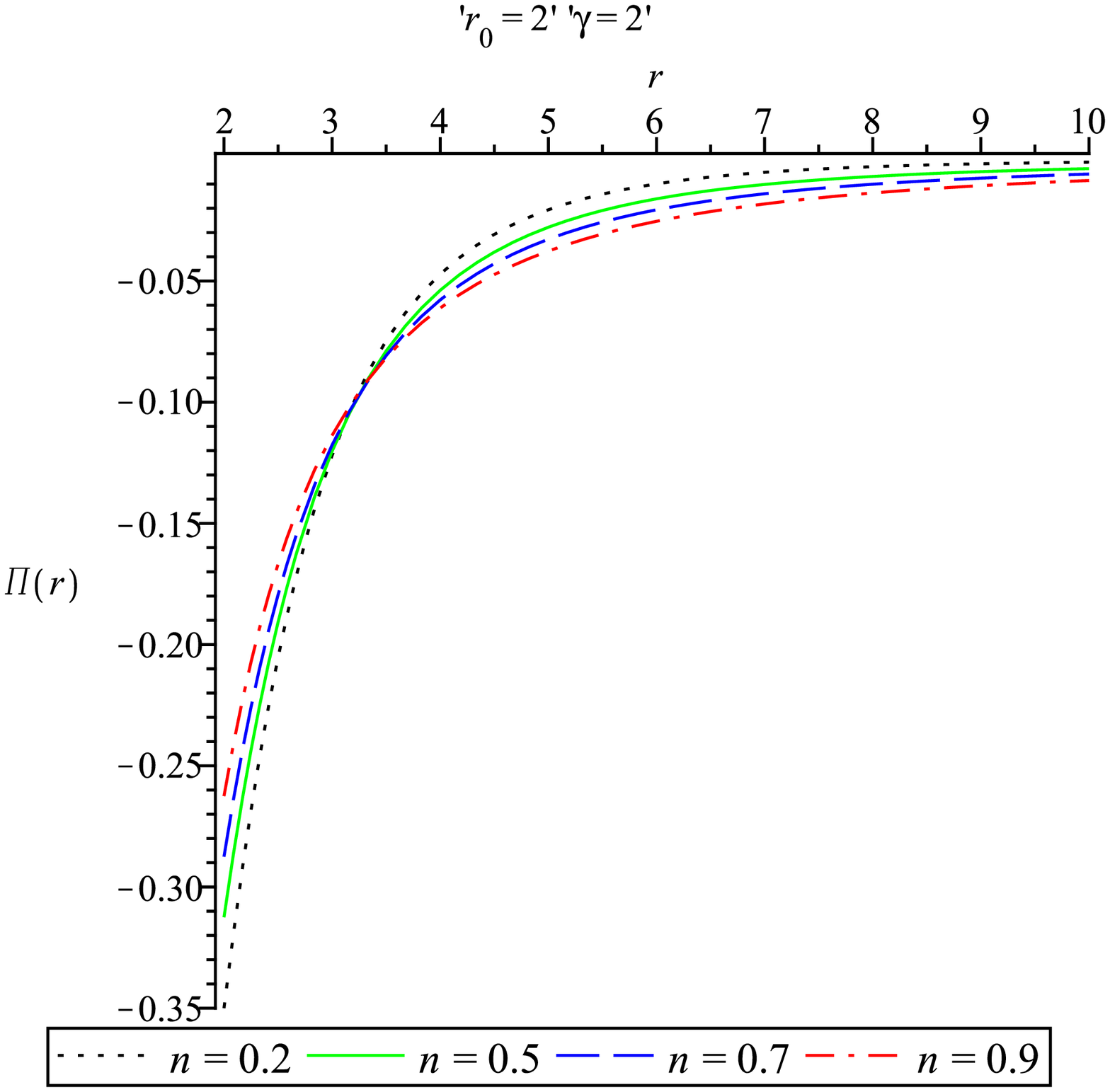}
\includegraphics[width=5.cm]{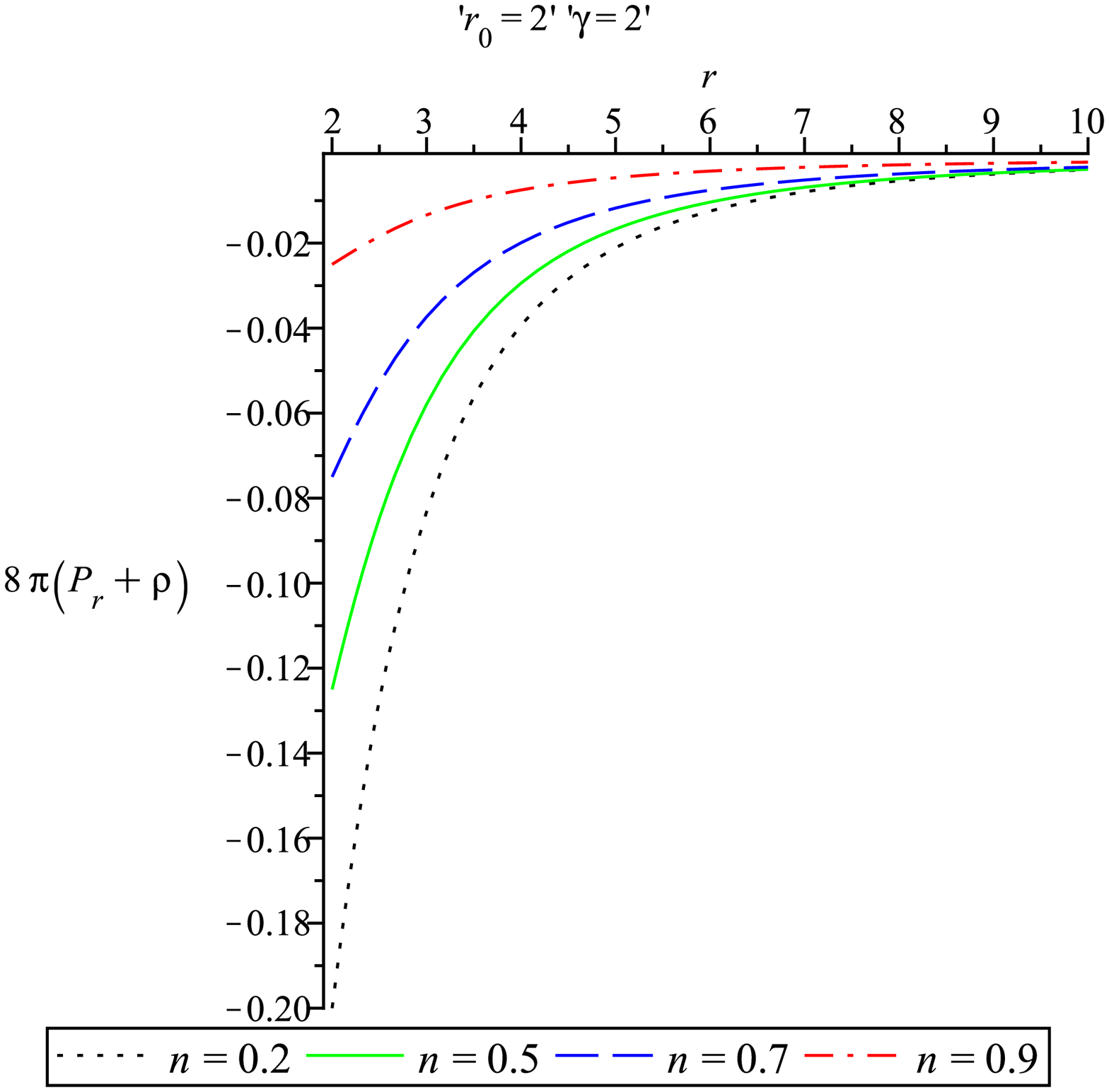}
\\
\end{tabular}
\end{center}
\caption{(left)Generating function $Z$ is always positive.    (middle) Generating function $\Pi$ is always negative.   (right) NEC  is violated .}
\end{figure*}

\begin{figure*}[thbp]
\begin{center}
\begin{tabular}{rl}
\includegraphics[width=4.5cm]{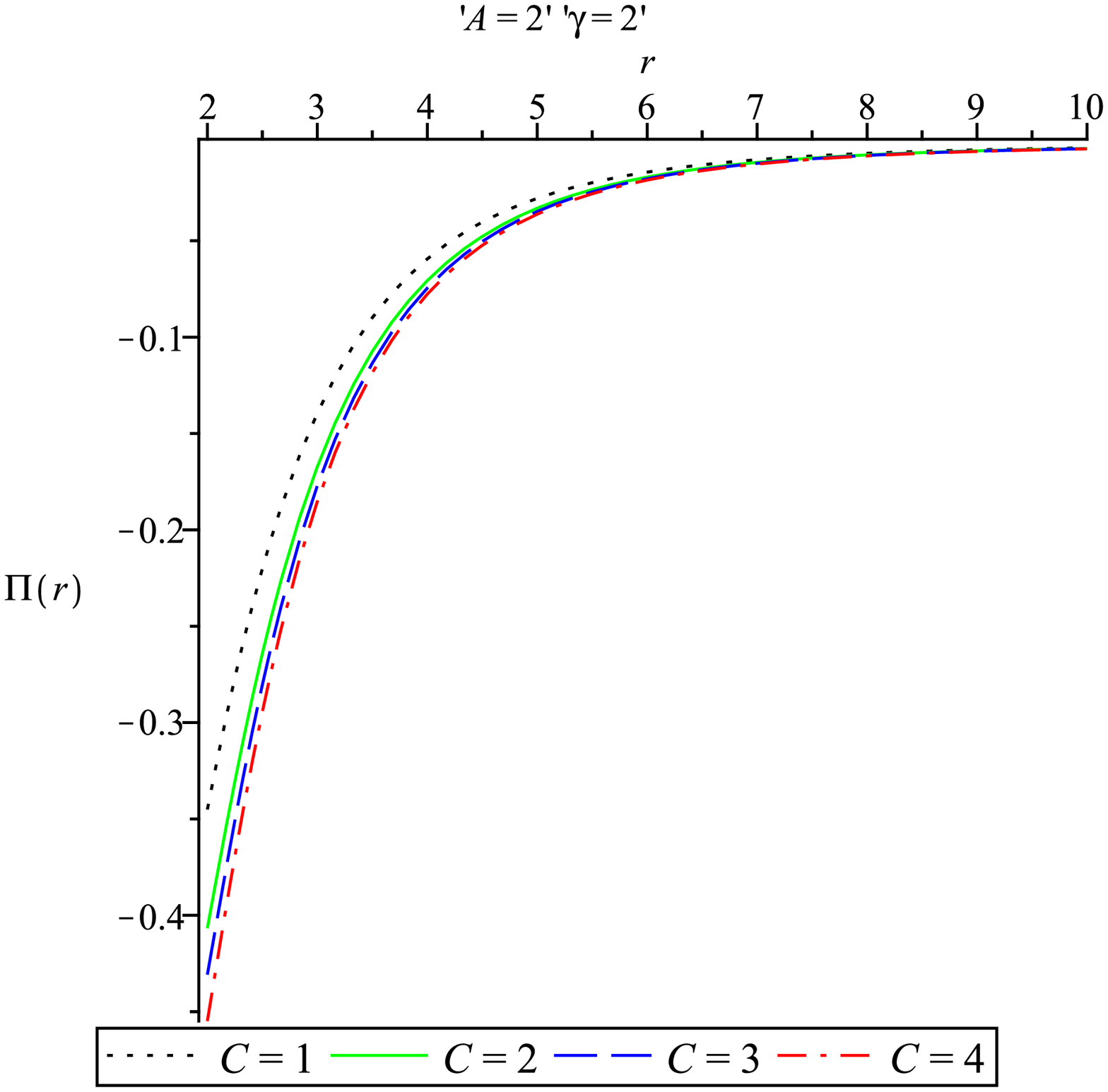}&
\includegraphics[width=4.5cm]{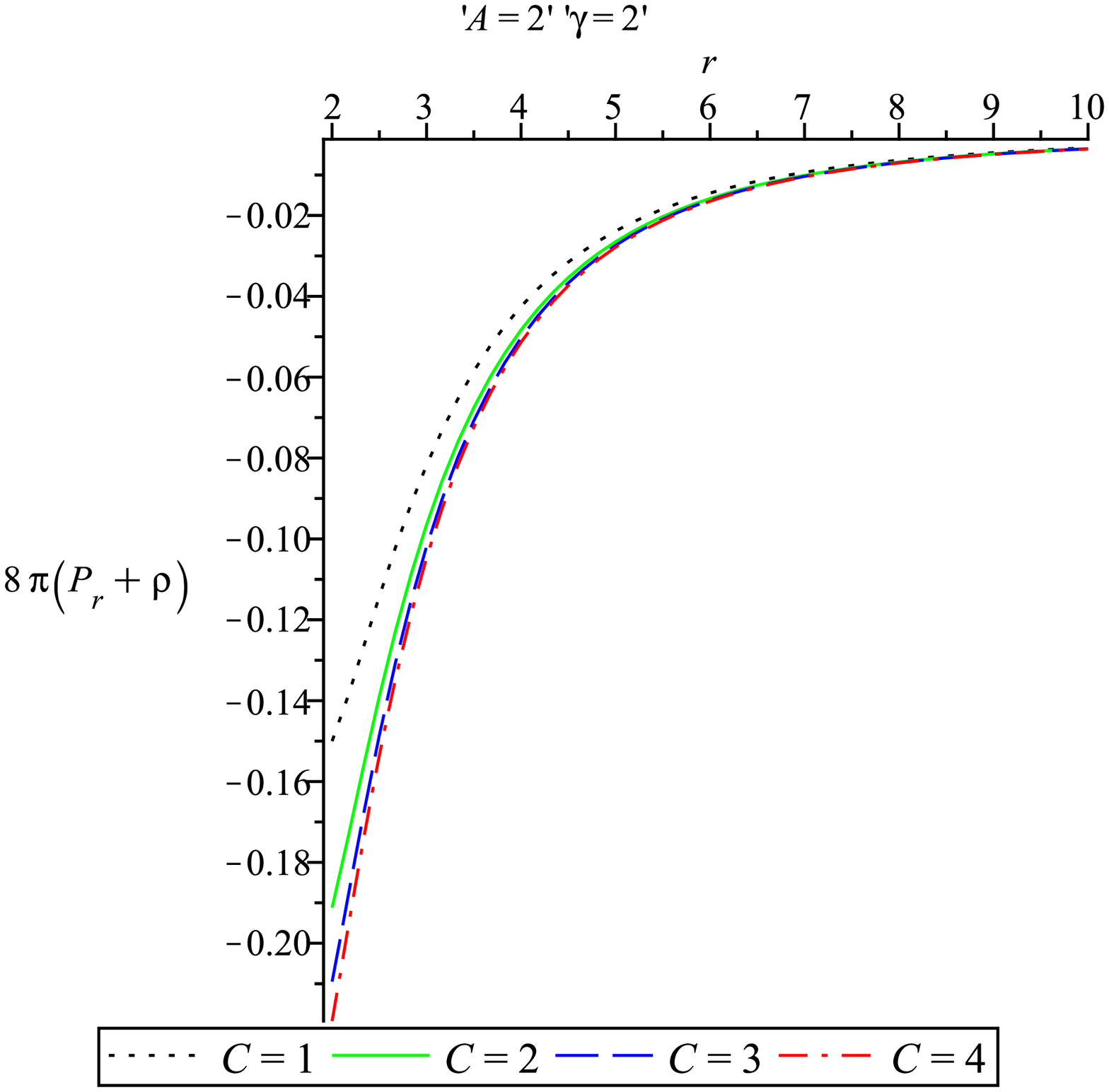}
\\
\end{tabular}
\end{center}
\caption{(left) Generating function $\Pi$ is always negative.   (right) NEC  is violated .}
\end{figure*}

 \begin{figure*}[thbp]
\begin{center}
\begin{tabular}{rl}
\includegraphics[width=4.5cm]{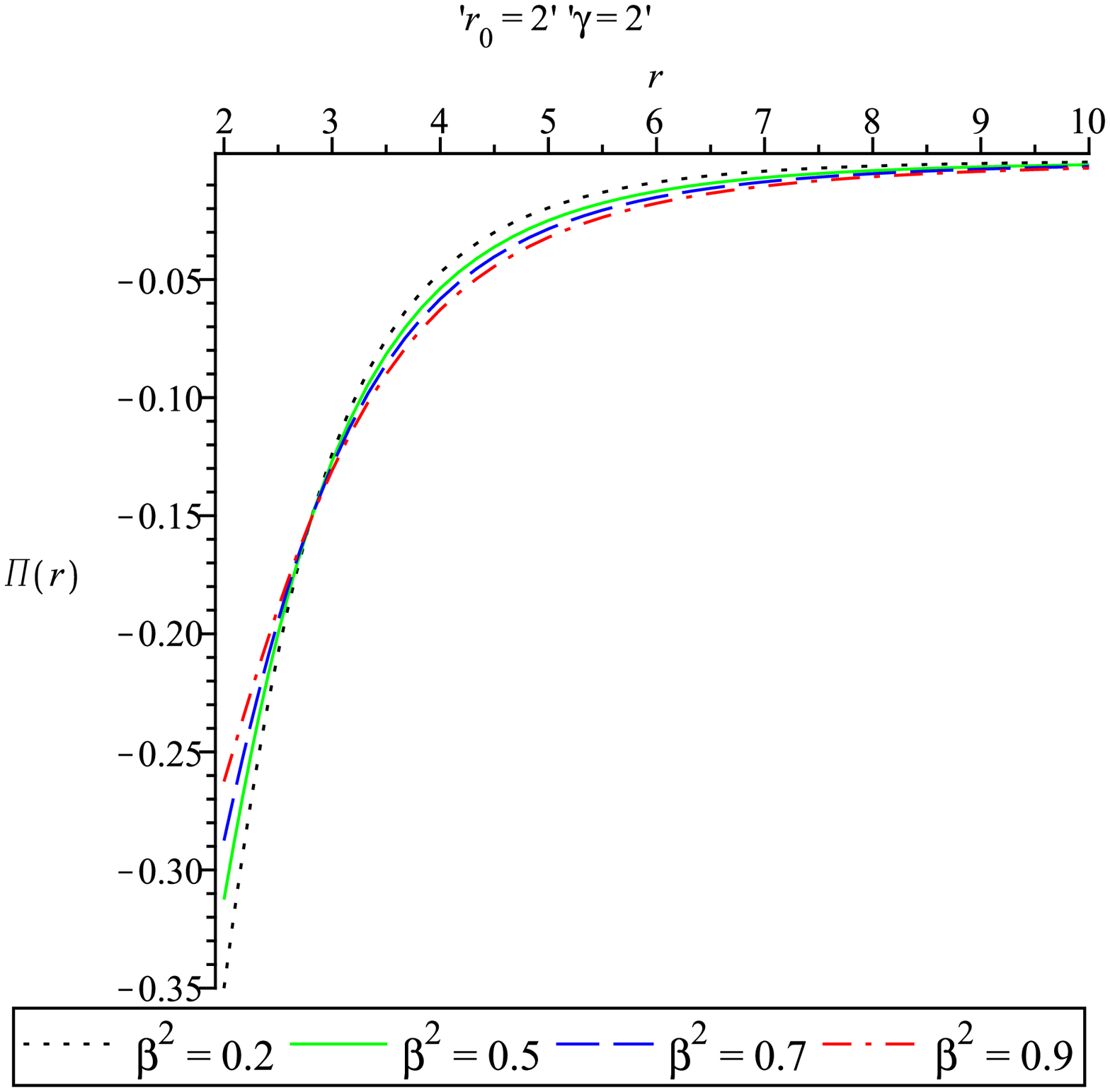}&
\includegraphics[width=4.5cm]{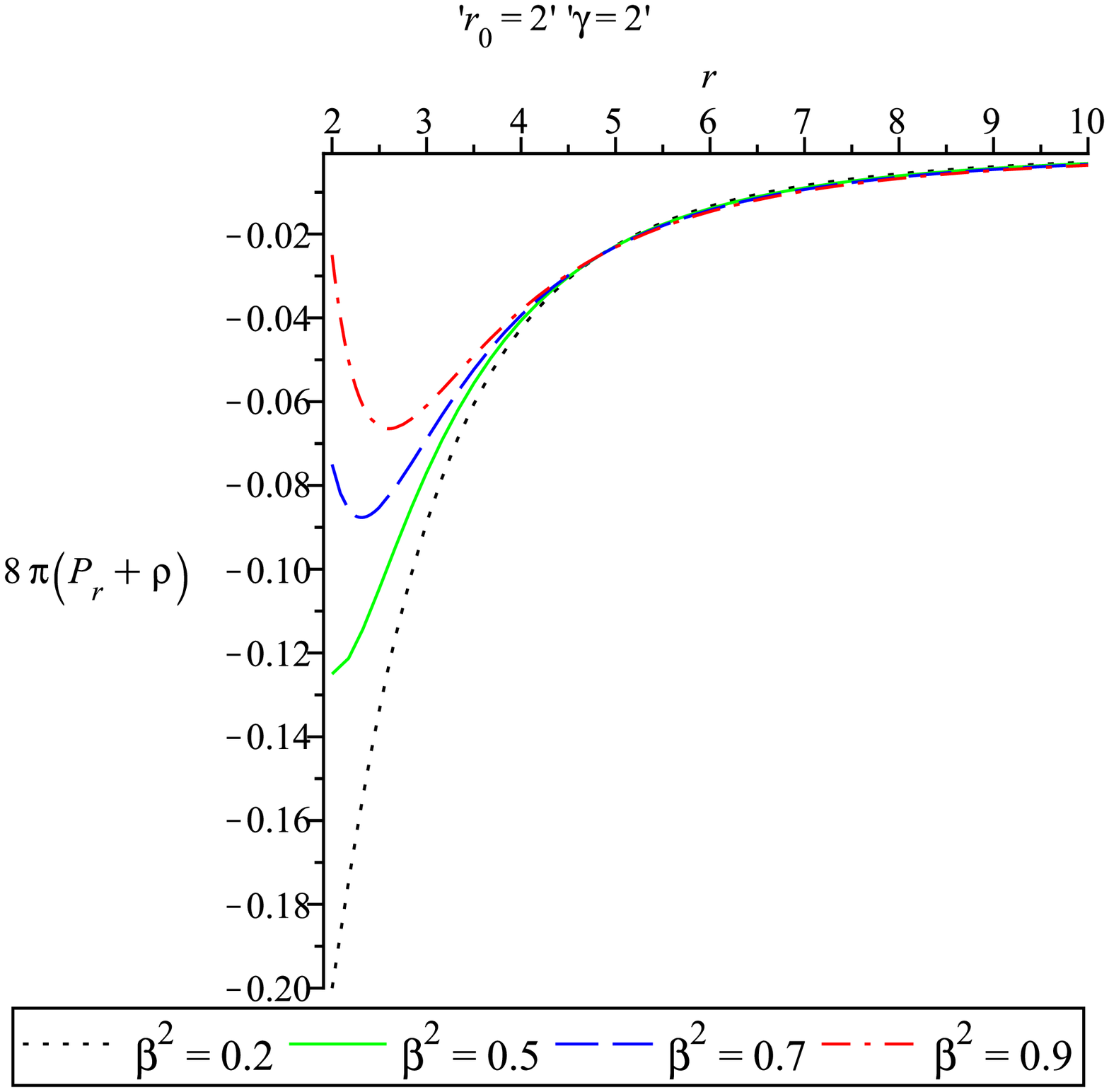}
\\
\end{tabular}
\end{center}
\caption{(left) Generating function $\Pi$ is always negative.   (right) NEC  is violated.}
\end{figure*}
We notice  that the   nature of the generating functions  matches with previous results and NEC is violated ( see figures 8 - 10 ).
\subsection{Specific generating function   and redshift function:}
In the previous sections,   wormhole spacetimes are created mainly by scheming a suitable
metric and followed by reconstructing the generating functions. Now, we will obtain some exact solutions    by assuming generating function and redshift functions.

 \subsubsection{   $f(r) = 0$:}
 Let us consider the specific generating function,  $\Pi(r)= - (\frac{a}{r^m}+hr^n)$, where, $ a, m, h,n$  are constants. In this case the equation (12) yields  the following  expression for shape function
 \begin{equation}
b(r)= \frac{a}{m-1}r^{3-m}-\frac{h}{n+1}r^{n+3}+C_3r^2,
\end{equation}
where $C_3$ is integration constant.

Note that if $C_3 \neq 0$ , $\frac{b(r)}{r} $ does not tend to zero for large $r$. In this case we can get  wormhole structure with finite size and we will have to use junction condition. However, for $C_3 =0 $ , we can get  usual  wormhole structure and we have checked it graphically ( see figure 11 ).  Observe that  we have throat as well flare-out condition is satisfied.

\begin{figure*}[thbp]
\begin{center}
\begin{tabular}{rl}
\includegraphics[width=5.cm]{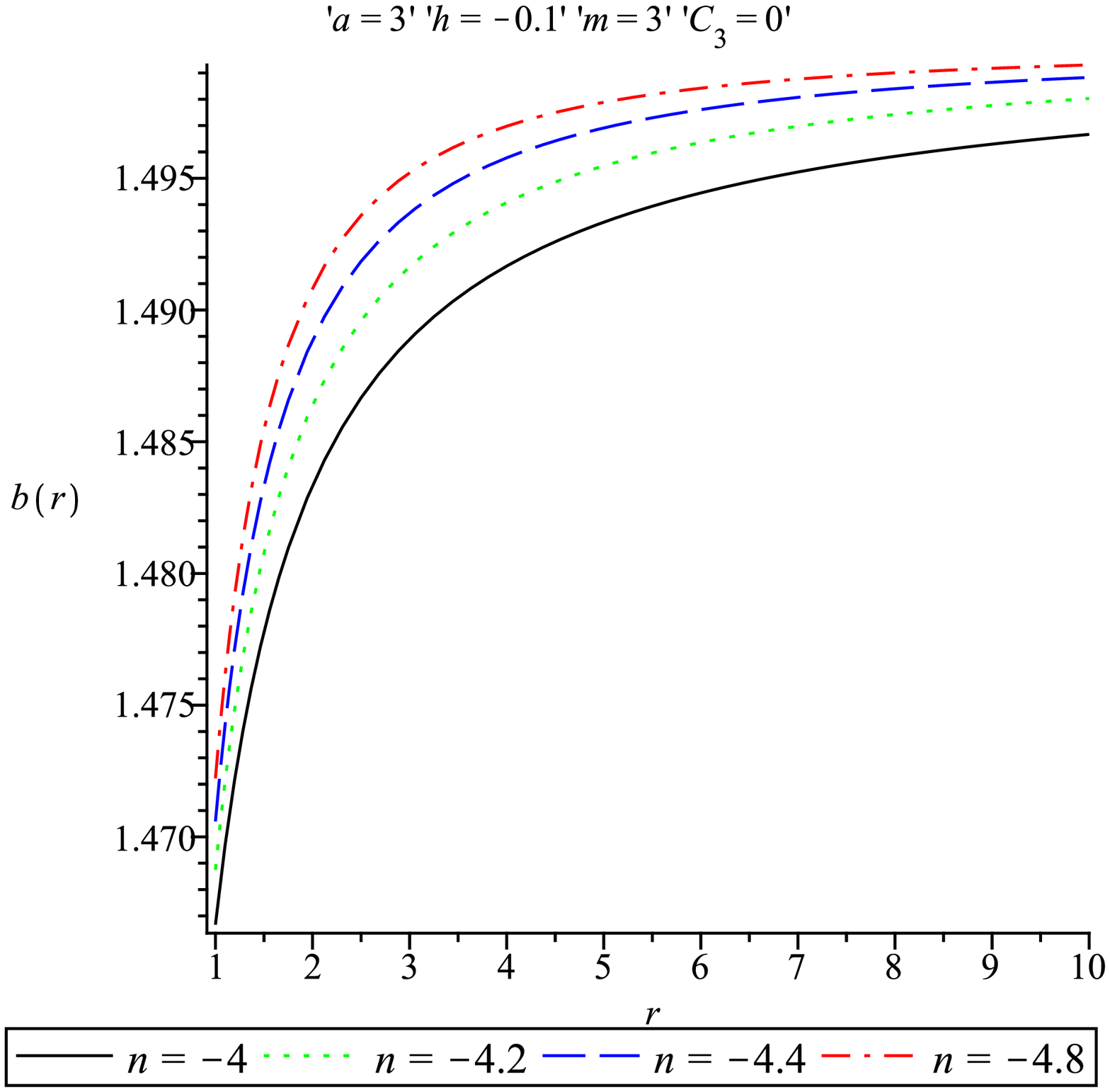}&
\includegraphics[width=5.cm]{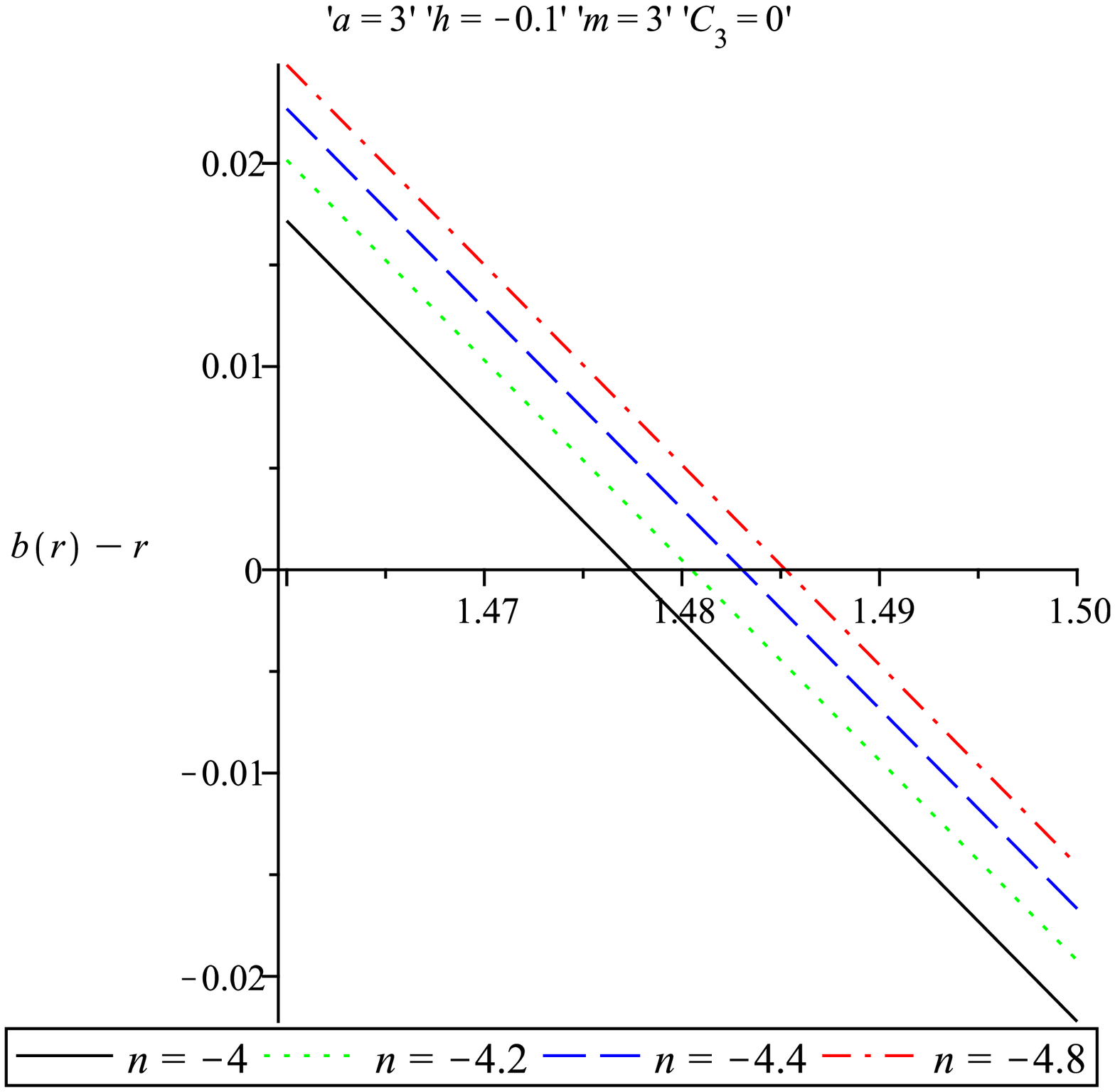}
\includegraphics[width=5.cm]{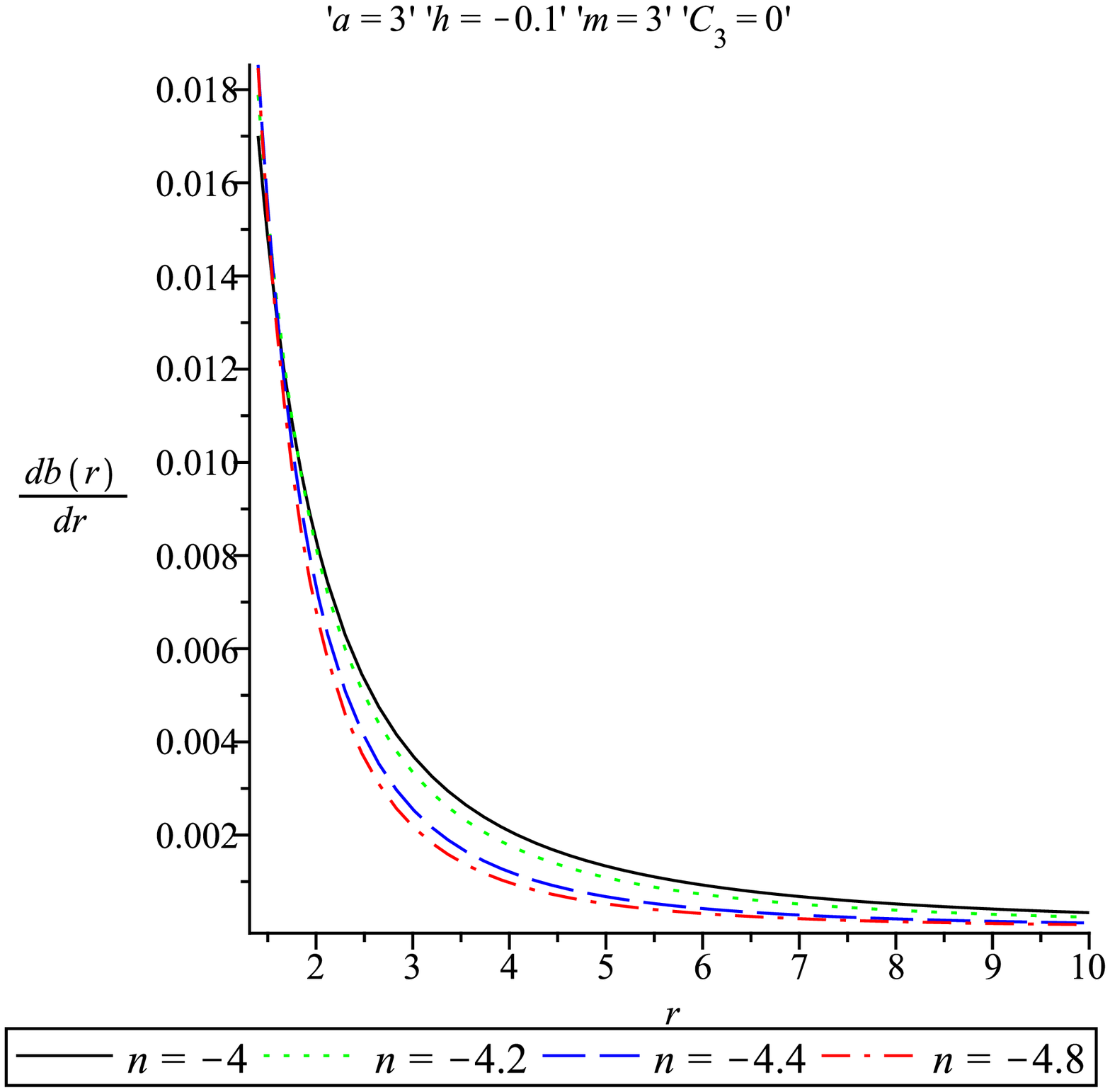}
\\
\end{tabular}
\end{center}
\caption{(left)Shape function of  wormhole. (middle) Throat of  wormhole is located where the $b(r)-r $ cuts $r$ axis.  (right) Flare-out condition  is satisfied .}
\end{figure*}

\begin{figure*}[thbp]
\begin{center}
\begin{tabular}{rl}
\includegraphics[width=4.5cm]{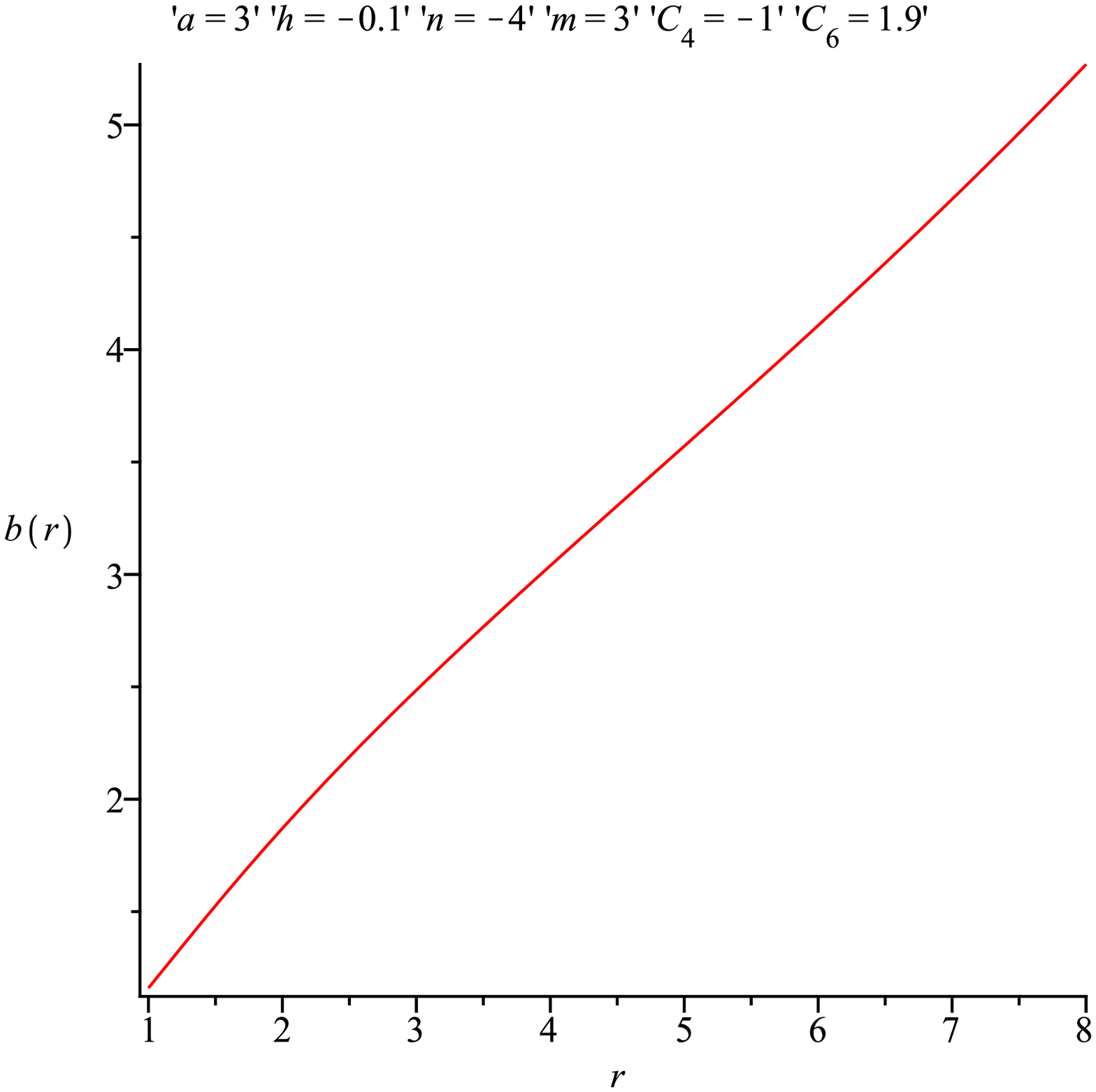}&
\includegraphics[width=4.5cm]{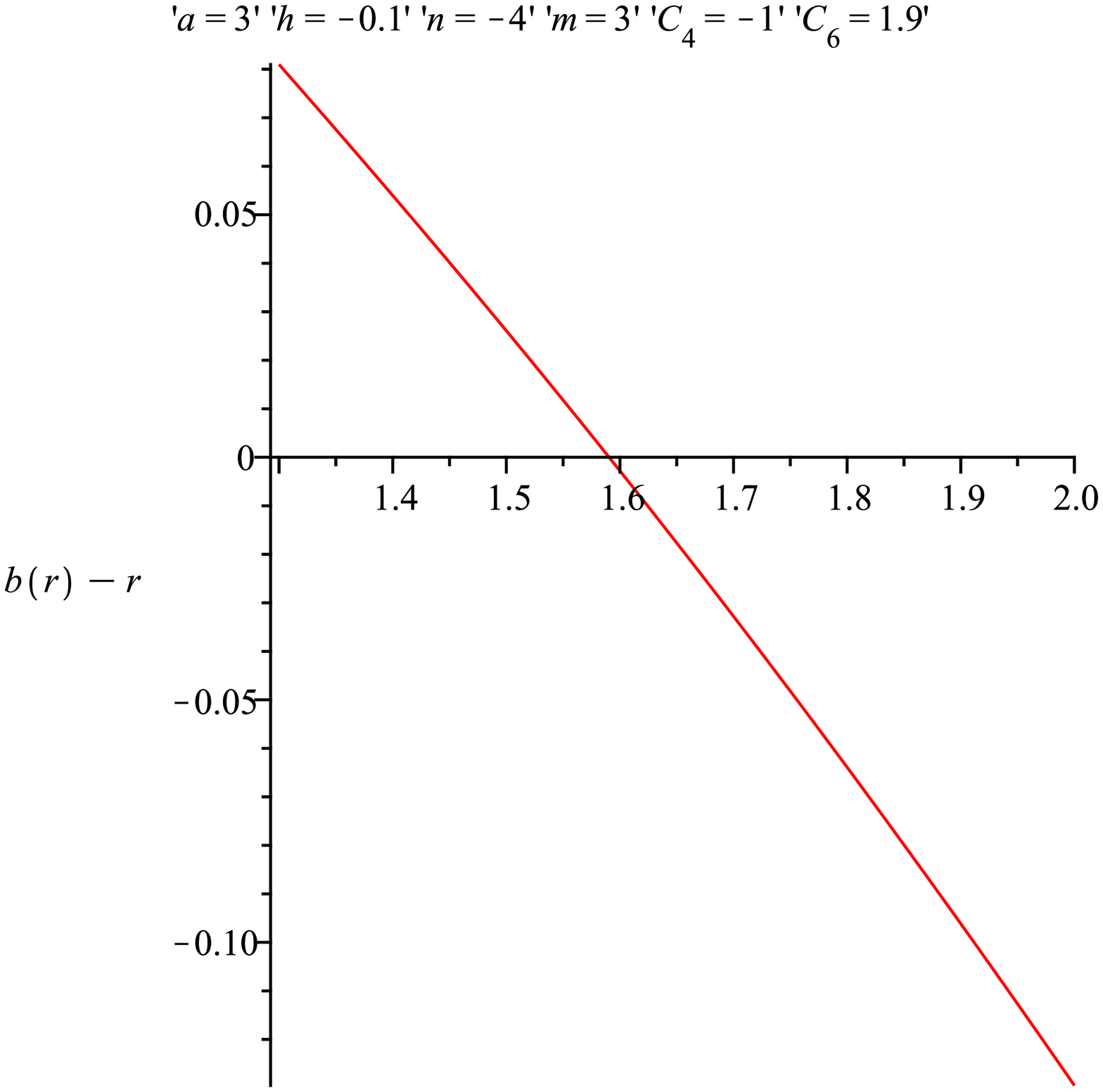}
\includegraphics[width=4.5cm]{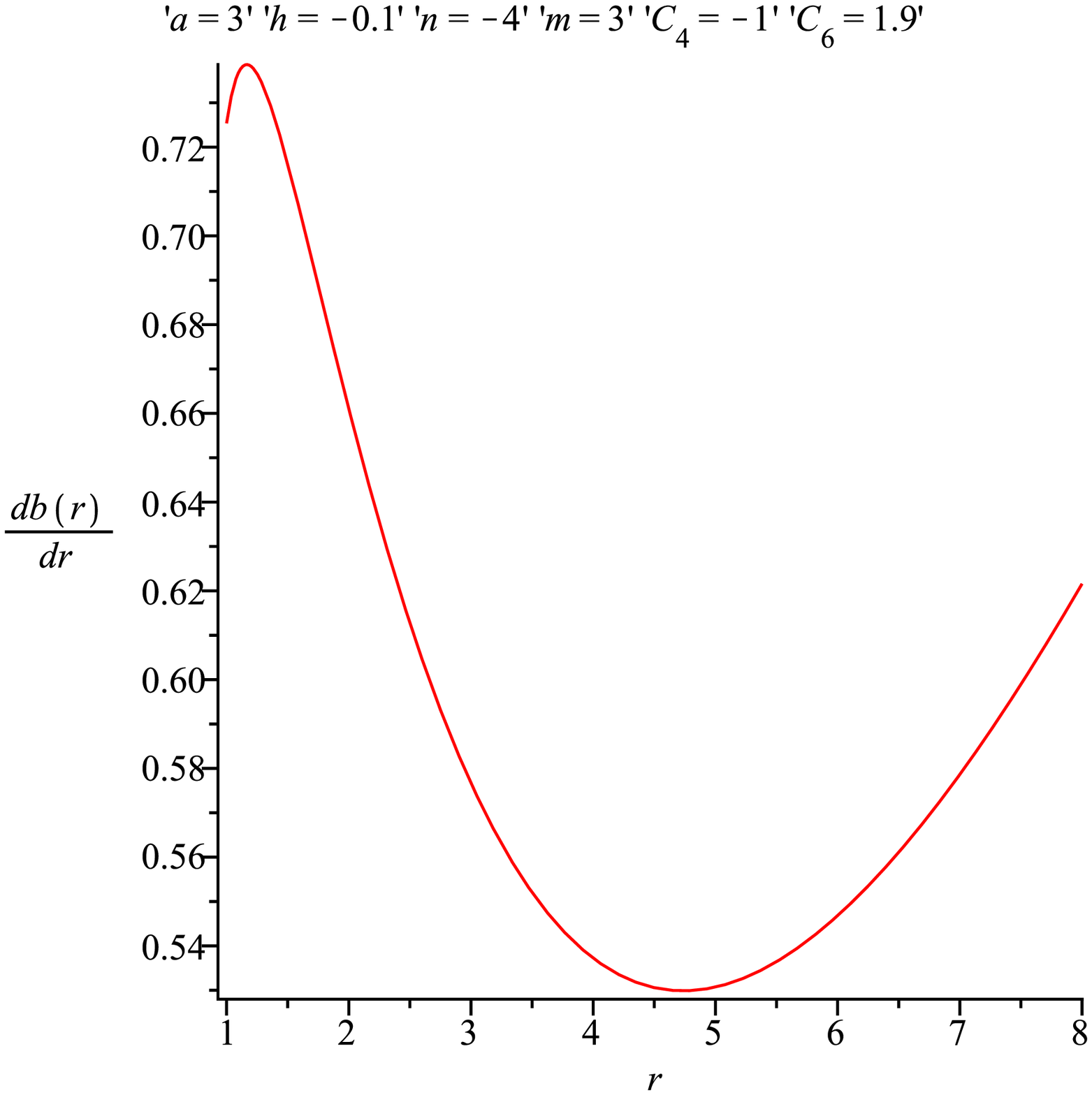}
\\
\end{tabular}
\end{center}
\caption{(left)Shape function of  wormhole  for   case 2.    (middle) Throat of  wormhole is located where the $b(r) -r $ cuts r axis.  (right) Flare-out condition  is satisfied .}
\end{figure*}

\subsubsection{Case  $f(r)=\frac{\alpha}{r}$:}
In this case the equation (12) yields a rather complicated expression. However, the shape function admits the following solution
\begin{equation}
b(r)= r-\frac{r^5}{C_4(r-\alpha)^3e^{\alpha/r}}\left[C_6-\int \frac{C_4(1-ar^{-m+2}-hr^{n+2})(r-\alpha)^2~e^{\alpha/r}}{r^4}dr\right ]
\end{equation}
where  $C_4$ and  $C_6$ are integration constant. Here the solution provides the  wormhole structure and we have checked it graphically ( see figure 12 ). However,  the  spacetime is not asymptotically flat, therefore,  it will have to be cut off at some  radial distance $r = R$ and  have to join to an external Schwarzschild spacetime.

\section{Concluding remarks}

In this work, for the first time, we have found the generating functions comprising the wormhole like geometry. Inspired by Herrera et al discovery of a new algorithm to obtain all static spherically symmetric describing anisotropic perfect fluid solutions,
we have considered in this work the spacetime metric of the wormhole and then deduce the generating functions. The specific solutions for the shape functions of the wormholes were obtained  by considering choices for the generating function and the redshift functions.
  In one  case, however, we see that the wormhole cannot be arbitrarily large and should be   cut off at some  radial distance $r = R$ and  have to join to an external Schwarzschild spacetime.
  In fact we have tried to find the nature of the generating functions for the wormhole by means of some examples.  As this is the first attempt, we have confined our self for particular examples. Our target was not merely to find some specific shape functions and redshift functions to find wormhole solutions . Rather, we try to investigate the nature of the generating functions, which is very much interesting to wormhole lovers. That means theoretically we can predict wormholes  by choosing generating functions.  Later we may examine the energy stress components.
  Note that we have found two generating functions : one ($Z(r)$) is related to the geometry of spacetime and other ($\Pi(r)$)  is related to the matter distribution comprising the wormhole. Interesting properties are found of generating functions as follow: the generating $Z(r)$ related to the redshift function  is always positive and decreasing function of radial coordinate and the second generating function ($\Pi(r)$) is always negative and increasing in nature. Both functions are asymptotic in nature for large $r$.  By monitoring the measure of anisotropy ($\Pi(r)$) of the matter distribution,  one can get   wormholes. That means one requires the knowledge of measure of anisotropy to generate some wormholes. However, assuming the metric functions and find the generating functions along with the study of physical fluid properties is easier than by assuming the fluid properties via anisotropy. Although, a wide variety of the solutions can be  found by choosing several generating functions, however, we emphasize that all solutions do not correspond   to wormhole physics. This will be investigated in a future work.

\section*{Acknowledgments}

FR would like to thank the authorities of the Inter-University Centre for Astronomy and Astrophysics, Pune, India for providing research facilities. FR and SS are also grateful to DST-SERB (Grant  No.: EMR/2016/000193)  and UGC (Grant  No.: 1162/(sc)(CSIR-UGC NET , DEC 2016),  Govt. of India,    for financial support respectively.


\begin{thebibliography}{77}

\bibitem[1]{1} Herrera et al , Phys.Rev.D 77, 027502 (2008)
\bibitem[2]{FL} Flamm L. Beitrage zur Einsteinschen Gravitationstheorie. Phys Z. 1916;17:448.
\bibitem[3]{ER} Einstein A, Rosen N., Phys Rev. 1935;48:73-7.
\bibitem[4]{WH} Wheeler J. A. Phys Rev. 1955;97:511-36
\bibitem[5]{VL} VisserM. Lorentzian wormholes: from Einstein to Hawking. New York: American Institute of Physics; 1995
\bibitem[6]{MW} Misner C. W., Wheeler J. A., Annals Phys. 1957;2:525
\bibitem[7]{whe} Wheeler J. A., Annals Phys. 1957; 2:604
\bibitem[8]{bro1} Bronnikov K. A., Acta Phys. Polonca 1973; B4:251
\bibitem[9]{ful} Fuller R. W., Wheeler J. A., Phys. Rev. 1962: 919:128
 \bibitem[10]{ell} Ellis H. G., J. Math. Phys. 1973; 14:104
\bibitem[11]{cle1} Clement G., Gen. Relativ. Grav., 1984; 16:131
\bibitem[12]{cle2} Clement G., Gen. Relativ. Grav., 1984; 16:447
\bibitem[13]{cle3} Clement G., Gen. Relativ. Grav., 1989; 21:849
\bibitem[14]{cle4} Clement G., Gen. Relativ. Grav., 1981; 13: 747
\bibitem[15]{MT}  M. S. Morris, K. S. Thorne, Am. J. Phys. 56 (1988) 395
\bibitem[16]{ML} M. Visser, Lorentzian wormholes: From Einstein to Hawking, (AIP Press, New York, 1995).
\bibitem[17]{8}S. Sushkov, Phys. Rev. D 71 (2005) 043520
\bibitem[18]{9}   M. Cataldo et al, Phys. Rev. D 79 (2009) 024005
\bibitem[19]{10} J.A. Gonzalez et al, Phys. Rev. D 79 (2009) 064027
\bibitem[20]{11} F. Rahaman et al, Phys.
Lett. B 633 (2006) 161
\bibitem[21]{12}  F. Rahaman et al, Phys. Scr. 76 (2007) 56
\bibitem[22]{13} P.K.F. Kuhfittig,
Class. Quantum Grav. 23 (2006) 5853
\bibitem[23]{14} F. Rahaman et al, Gen.Rel.Grav. 39 (2007) 145
\bibitem[24]{15}   F.S.N. Lobo, Phys. Rev. D 71 (2005) 084011
\bibitem[25]{16}  Rahaman F. et al., Eur. Phys. J. C. 74, 2750 (2014)
\bibitem[26]{17}Rahaman F. et al., Annals of Physics, 350, 561567 (2014)
\bibitem[27]{18} Farook Rahaman, G.C. Shit, Banashree Sen  , Saibal Ray
, Astrophys.Space Sci. 361 , 37 (2016)
\bibitem[28]{19} F. Rahaman, M. Kalam, M. Sarker, A. Ghosh, B. Raychaudhuri, Gen. Relativ. Gravit. 39, 145 (2007)
\bibitem[29]{20}  F Rahaman et al, Int.J.Theor.Phys. 48 (2009) 471-475
\bibitem[30]{22} F Lobo, Phys.Rev. D73 (2006) 064028
\bibitem[31]{21}  F Rahaman et al, Int.J.Theor.Phys. 48 (2009) 1637-1648
\bibitem[32]{bro5} Bronnikov K. A., Chervon S. V., Sushkov S. V., Grav. Cosmo. 2009: 15:241
\bibitem[33]{bro2} Bronnikov, K. A., Sushkov, S., Class. Quantum Grav. 2010: 27:095022
\bibitem[34]{bro3} Bronnikov, K. A., Baleevskikh K.A., Skvortsova M. V., Phys. Rev. D 2017; 96: 124039








\end{thebibliography}
\end{document}